%% file: master.tex
\documentclass[aps,prx,twocolumn,superscriptaddress]{revtex4-2}
\usepackage{amsmath,amssymb}
\usepackage{mathrsfs}
\usepackage{graphicx}
\usepackage{setspace}
\usepackage{braket}
\usepackage{dcolumn}
\usepackage{bm}
\usepackage{hyperref}
\usepackage{cleveref}
\usepackage{physics}
\usepackage{xcolor}
\DeclareMathOperator{\sinc}{sinc}

\begin{document}

\input{main}

\clearpage
\onecolumngrid

\setcounter{section}{0}
\setcounter{equation}{0}
\setcounter{figure}{0}
\setcounter{table}{0}

\renewcommand{\thesection}{S\arabic{section}}
\renewcommand{\theequation}{S\arabic{equation}}
\renewcommand{\thefigure}{S\arabic{figure}}
\renewcommand{\thetable}{S\arabic{table}}

\begin{center}
{\Large \bf Supplementary Information}
\end{center}

\input{supplement}

\end{document}

%% file: main.tex
\title{Quantum computing and quantum optics with recoiled free electrons}

\author{Maxim Sirotin}
\email{msirotin@g.harvard.edu}
\thanks{Corresponding author}
\affiliation{Department of Physics, Friedrich-Alexander-Universit\"at Erlangen-N\"urnberg, 91058 Erlangen, Germany}
\affiliation{Max Planck Institute for the Science of Light, 91058 Erlangen, Germany}
\affiliation{Department of Physics, Harvard University, Cambridge, MA 02138, USA}
\affiliation{Research Laboratory of Electronics, Massachusetts Institute of Technology (MIT), Cambridge, MA 02139, USA}

\author{Andrei Rasputnyi}
\email{andrei.rasputnyi@mpl.mpg.de}
\affiliation{Department of Physics, Friedrich-Alexander-Universit\"at Erlangen-N\"urnberg, 91058 Erlangen, Germany}
\affiliation{Max Planck Institute for the Science of Light, 91058 Erlangen, Germany}

\author{Tom\'a\v{s} Chlouba}
\email{chloubatom@gmail.com}
\affiliation{Department of Physics, Friedrich-Alexander-Universit\"at Erlangen-N\"urnberg, 91058 Erlangen, Germany}

\author{Roy Shiloh}
\email{roy.shiloh@mail.huji.ac.il}
\affiliation{Department of Physics, Friedrich-Alexander-Universit\"at Erlangen-N\"urnberg, 91058 Erlangen, Germany}
\affiliation{Institute of Applied Physics, Hebrew University of Jerusalem, Jerusalem 9190401, Israel}

\author{Peter Hommelhoff}
\email{peter.hommelhoff@fau.de}
\thanks{Corresponding author}
\affiliation{Department of Physics, Friedrich-Alexander-Universit\"at Erlangen-N\"urnberg, 91058 Erlangen, Germany}
\affiliation{Faculty of Physics, Ludwig-Maximlians-Universit\"at M\"unchen, 80539 M\"unchen, Germany}

\begin{abstract}
Free electrons interacting coherently with optical fields provide a powerful platform for quantum simulation and quantum control. For kiloelectron-volt electron energies, even optical photon emission and absorption produce appreciable quantum recoils, endowing the electron with a discrete and controllable energy ladder. Starting from relativistic quantum electrodynamics, we derive an exact recoil-resolved interaction Hamiltonian in a traveling wave picture. The resulting recoil ladder forms a high-dimensional qudit with programmable couplings and sufficient controllability for universal quantum computation. We demonstrate applications to quantum simulation, including one-dimensional analogue black-hole models including Hawking radiation physics, and to quantum information processing, where multiple logical qubits and high-fidelity gates can be realized with a single electron. In parallel, the same recoil-enabled dynamics enable the controlled creation of complex hybrid electron--photon states, in which engineered ladder transitions imprint nonclassical correlations and structure onto the emitted light. Together, these results establish recoiled free electrons as a versatile platform bridging quantum optics, Hamiltonian engineering, and quantum simulation.
\end{abstract}

\maketitle

\section{Introduction}\label{sec:intro}

Free-electron quantum optics provides a powerful framework for exploring quantum light--matter interactions beyond bound electronic systems \cite{ruimy2025free, de2025roadmap}. In this approach, swift electrons—most prominently relativistic electrons in transmission electron microscopes (TEMs)—are treated as coherent quantum waves that exchange energy and momentum with optical and electromagnetic fields. Early demonstrations of photon-induced near-field electron microscopy (PINEM) showed that free electrons can undergo coherent, phase-controlled interactions with optical fields, leading to discrete energy sidebands and optical control of electron wavefunctions \cite{barwick_photon-induced_2009}. Subsequent work established quantum-coherent electron--photon coupling in TEMs \cite{feist_quantum_2015}, enabling fully quantized descriptions of electron--light interactions, cavity-mediated coupling, and electron--photon entanglement \cite{feist_cavity-mediated_2022, huang_electron-photon_2023}. 

These developments have enabled advances in ultrafast imaging \cite{fishman_imaging_2023, gaida_lorentz_2023, bucher_coherently_2023, chirita2025light}, electron acceleration \cite{england_dielectric_2014, adiv_quantum_2021, shiloh_miniature_2022}, and hybrid light--matter physics \cite{meier_few-electron_2023, Haindl2023, heimerl_multiphoton_2024, heimerl2025quantum}. In most of these existing applications, electrons propagate in the (near-)relativistic regime, where the large electron momentum strongly suppresses quantum recoil---that is, changes to the momentum-conservation (phase-matching) condition following photon emission or absorption. As a result, free-electron quantum optics has largely been explored in a recoil-free limit.

At lower electron energies in the kiloelectron-volt range, accessible in scanning electron microscopes (SEMs), the role of recoil changes qualitatively. Free electrons can still be phase-matched to optical photons, enabling strong and programmable coupling to nanophotonic and cavity structures, but photon absorption and emission now induce a measurable change in the electron momentum \cite{huang_quantum_2023, karnieli_jaynes-cummings_2023, karnieli_universal_2023, karnieli2024strong}. This recoil gives rise to a discrete, individually addressable ladder of electron energy states while preserving a large accessible Hilbert space. SEMs provide an attractive experimental environment for exploring this regime, offering large vacuum chambers, excellent optical access, and flexible integration with photonic structures. Recent demonstrations of PINEM and nanophotonic acceleration in SEMs highlight the feasibility and promise of this platform \cite{shiloh_quantum-coherent_2022, chlouba_coherent_2023, sirotin_tunable_2023}.

Rather than treating recoil as a limitation, it can be harnessed as a resource for quantum control. The recoil-induced energy ladder transforms the free electron from a passive probe into an active quantum system with a controllable internal structure in the sense of a shaped matter wave. The ladder forms an anharmonic oscillator with programmable nearest-neighbor couplings, constituting a high-dimensional ladder qudit with full and universal control. This ladder qudit enables two complementary capabilities. First, it provides a natural platform for quantum simulation, where programmable couplings realize effective lattice Hamiltonians. Using globally applied optical tones, by which we mean externally applied laser fields with specified optical frequencies, phases, and amplitudes, together with momentum-mismatch-induced selectivity, we show how one-dimensional analogue black-hole Hamiltonians can be implemented, allowing simulations of curved spacetime and Hawking-like radiation physics within a single propagating electron. Second, the same ladder qudit supports quantum information processing: multiple logical qubits can be encoded within a single electron, and universal qudit and encoded-qubit gates can be realized using ultrafast optical control.

The same recoil-enabled dynamics also facilitate the controlled generation of nonclassical quantum states of light and hybrid light--matter states, including single-photon \cite{degen_quantum_2017}, GKP \cite{dahan_creation_2023}, GHZ \cite{greenberger_bells_1990}, NOON, and squeezed-vacuum states \cite{mckenzie_squeezing_2004, grote_first_2013, zhong_phase-programmable_2021}. Coherent energy and momentum exchange entangles the electron’s ladder degrees of freedom with photonic modes, imprinting engineered quantum correlations onto the emitted light. Programmable ladder couplings allow shaping of the joint electron--photon state, enabling heralded photonic states, correlated photon emission, and structured multimode light that are difficult to generate using purely optical nonlinearities.

Starting from first principles within a relativistic quantum electrodynamics framework \cite{cohen-tannoudji_photons_1997}, we derive the exact interaction Hamiltonian governing the coherent coupling between free electrons and optical fields in the presence of recoil. This approach allows us to identify effective Hamiltonians describing recoil-induced ladders, programmable couplings, and non-autonomous dynamics, and to obtain exact analytical solutions in experimentally relevant limits. Together, these results establish recoiled free electrons as a powerful platform bridging quantum optics, Hamiltonian engineering, and quantum simulation with electron microscopy and quantum-enhanced sensing, enabling controlled quantum dynamics in hybrid light--matter systems.

\section{Quantum simulations with recoiled free electrons}\label{sec:QC}

\begin{figure*}[ht]
\centering
\includegraphics[width=1\linewidth]{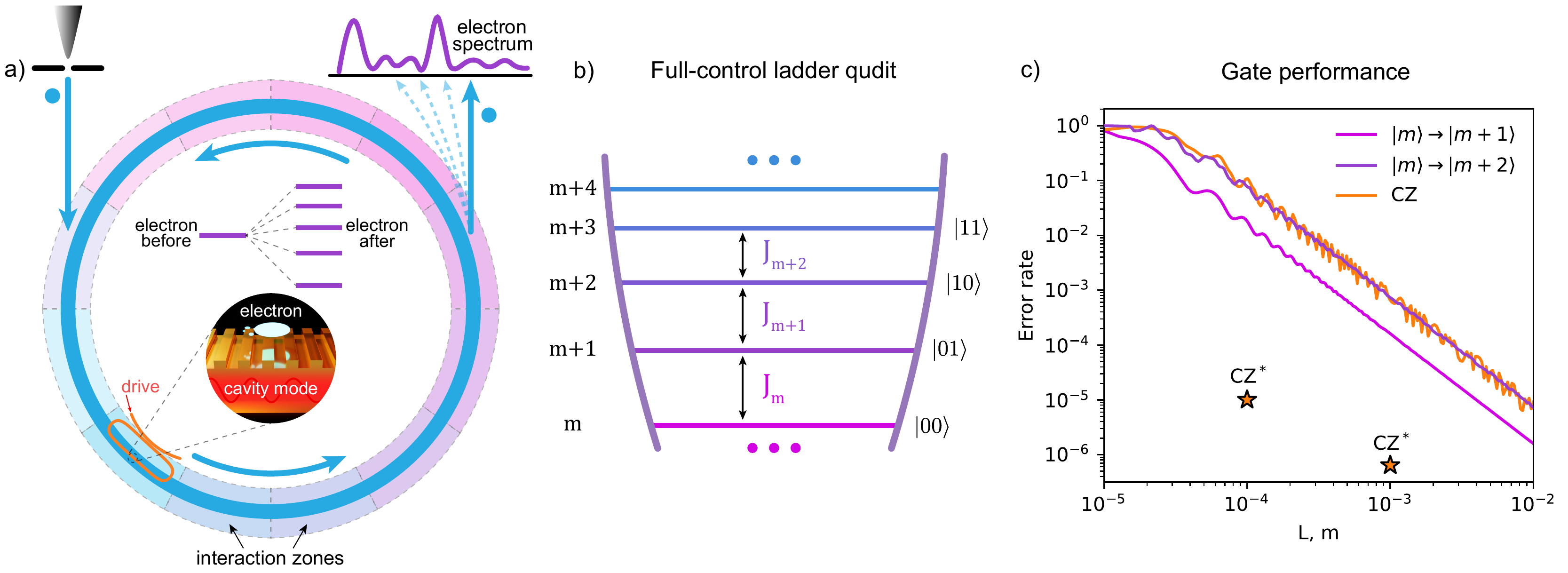}
\caption{\label{fig:QC} Quantum simulations with a recoiled free electron.
{a)} Conceptual schematic of an electron undergoing cyclotron motion and sequentially interacting with multiple cavities. Inset: electron--cavity interactions generate a synthetic anharmonic electron energy ladder, whose spectrum encodes the simulation outcome.
{b)} The anharmonic ladder functions as a universally controllable qudit.
{c)} Two qubits are encoded in four ladder levels, \(\ket{m}\)--\(\ket{m+3}\), and gate performance (SWAP and controlled-phase, CZ) is shown as a function of interaction length per zone. See text for details.}
\end{figure*}

We consider the interaction of free electrons with the evanescent electric field of optical cavity structures, as illustrated schematically in Fig.~\ref{fig:QC}. Free electrons generated in a transmission or scanning electron microscope (TEM or SEM), or in on-chip electron sources, are prepared at the desired kinetic energy and brought into close proximity with nanophotonic or cavity-enhanced near fields. In the presence of externally applied optical drives, the cavity supports multiple modes and optical tones, enabling strong and programmable electron--photon coupling. As the electron propagates past the cavity, it can coherently exchange energy and momentum with the optical field through near-field Cherenkov radiation \cite{cerenkov_visible_1934, tamm1937coherent, ginzburg1940quantum, tamm_general_1960, ginzburg1996radiation, adiv_observation_2023}, or, in the presence of a diffraction grating, through Smith--Purcell-type coupling into cavity modes \cite{SmithPurcell1953,doucas_first_1992, remez_spectral_2017, konakhovych2021internal}. Throughout this section, we focus on the {driven} regime, in which the optical field is pumped and composed of multiple controllable tones. In this limit, the interaction is dominated by stimulated processes, and if a cavity mode is not pumped at a given tone, the corresponding spontaneous contribution can be neglected. This setting enables programmable Hamiltonian engineering and the realization of effective lattice models, including curved-spacetime and analogue black-hole simulations, within a single propagating electron.

In the recoiled free-electron regime, quantum recoil---namely the change of electron momentum following photon emission or absorption---plays a central role in the dynamics. Because momentum conservation (phase matching) must be satisfied at each interaction step, the momentum change of the electron after emitting/absorbing a photon generally forces any subsequent photon to have a different frequency. As a result, if the cavity does not support additional modes or is not driven at the corresponding frequencies, further emission is suppressed, whereas in a driven multimode cavity the electron can continue emitting photons into modes of different frequencies. In this sense, quantum recoil acts as an effective {spectral filter} that selects the allowed transitions (see Supplementary Information \Cref{sec:Supp_recoil} for details).

At electron energies in the kiloelectron-volt range, the longitudinal motion of a free electron can be phase-matched to optical photons, enabling coherent light--matter interactions while simultaneously producing resolvable recoil. This recoil-induced selectivity gives rise to a discrete ladder of electron energy states that are coherently addressable and can be coupled in a programmable manner, forming an effective synthetic lattice with a large accessible Hilbert space. The combination of strong, momentum-selective coupling and free-space propagation makes this regime particularly well suited for quantum simulation of engineered lattice Hamiltonians. Experimentally, these conditions are naturally realized in scanning electron microscopes and related electron-optical platforms, which provide long interaction lengths, excellent optical access, and flexible integration of photonic structures.

Starting from the relativistic Dirac field of the free electron coupled to the electromagnetic field and transforming to a length-propagation picture (see Supplementary Information \Cref{sec:Supp_hamiltonian}), we obtain an exact recoil-resolved description in a basis of energy sidebands \(\{\ket{m}\}\), where \(\ket{m}\) denotes the electron after exchanging \(m\) photons with the optical field. The propagation coordinate \(z\) along the electron trajectory plays the role of the evolution parameter, such that the electron state \(\ket{\psi(z)}\) obeys a Schr\"odinger equation in \(z\)  (with \(\hbar = 1 \)),
\begin{equation}
i\,\frac{d}{dz}\ket{\psi(z)} = H(z)\ket{\psi(z)},
\end{equation}
so hereafter, ``non-autonomous'' refers to explicit dependence on the evolution coordinate \(z\), rather than on laboratory time.

In the driven (pumped) regime relevant for quantum simulation, interaction with externally applied optical tones yields an explicitly non-autonomous Hamiltonian,
\begin{equation}
H(z)
=
\sum_{m,j}
\left[
J_j\,
e^{i\left(k_{m+1}-k_m-\varkappa_j\right)z}\,
\ket{m{+}1}\!\bra{m}
+
\mathrm{h.c.}
\right],
\label{eq:H_nonauto}
\end{equation}
where \(k_m\) is the longitudinal electron wavevector of ladder state \(\ket{m}\), \(\varkappa_j\) is the wavevector of the \(j\)th applied optical tone, and \(J_j\) is the effective complex coupling coefficient induced by the \(j\)th optical tone. Each tone couples globally to all ladder transitions, while selectivity arises from the momentum mismatch \(k_{m+1}-k_m-\varkappa_j\), which produces rapidly oscillating phase factors along the propagation direction \(z\).

The coupling coefficient \(J_j\) encapsulates both the intrinsic vacuum electron--photon coupling strength and the amplitude of the applied optical field. Microscopically, one may write
\begin{equation}
J_j \;\propto\; g_{\rm Qu}\,\mathcal{E}_j,
\end{equation}
where \(g_{\rm Qu}\) is the vacuum coupling strength \cite{kfir_entanglements_2019,ZhaoUpperBound_g_quant} and \(\mathcal{E}_j\) is the complex field amplitude of the \(j\)th tone at the electron trajectory. In the absence of an external drive, the dynamics reduce to the fully quantized vacuum-coupling regime discussed in the quantum state generation sections below.

In the high-recoil regime, where \(|k_{m+1}-k_m-\varkappa_j|L\gg1\) for all off-resonant transitions over an interaction length \(L\), the rapidly oscillating off-resonant contributions average to zero. The dynamics then reduce to an effective autonomous tight-binding Hamiltonian,
\begin{equation}
H_{\mathrm{eff}}
=
\sum_m
J_m
\left(
\ket{m{+}1}\!\bra{m}
+
\ket{m}\!\bra{m{+}1}
\right),
\label{eq:H_eff}
\end{equation}
with programmable nearest-neighbor couplings \(J_m\) set by the chosen tones and their phases. As shown in the Supplementary Information \Cref{sec:Supp_lattice} and \Cref{sec:Supp_universality}, independent control of these couplings implies universal controllability on any finite ladder truncation, both in the qudit sense and in the Lloyd--Braunstein continuous-variable universality framework \cite{lloyd1999quantum, eriksson2024universal}.

Figure~\ref{fig:QC} summarizes the key elements of the recoiled free-electron platform and illustrates its capabilities for quantum simulation and quantum information processing. The system is based on a free electron propagating through a sequence of engineered interaction regions, where coherent coupling to optical cavity modes induces controlled energy exchange via photon recoil.

Figure~\ref{fig:QC}a shows a conceptual schematic of the platform. An electron sequentially interacts with multiple optical cavities, each defining a programmable interaction zone, corresponding to a spatial segment with independently programmable optical coupling. Photon absorption and emission induce discrete shifts in the electron’s longitudinal energy, generating a synthetic anharmonic ladder of recoil sidebands. Depending on the target operation, elementary transformations---such as nearest-neighbor swaps or phase shifts---can be implemented within a single interaction zone, while more complex quantum simulations or multiqubit gate sequences require a series of interaction zones applied in succession.

These interaction zones can be arranged linearly along the electron trajectory or, more conveniently, placed around a closed orbit using cyclotron motion. In such a configuration, the electron repeatedly traverses a circular path and encounters the same or different interaction zones on each revolution, enabling scalable and optically reconfigurable dynamics in a compact geometry. For example, a \(1~\mathrm{keV}\) electron has a velocity of \(v\simeq 0.06c\), so a cyclotron orbit with diameter \(10~\mathrm{cm}\) corresponds to a revolution period of \(\simeq 17~\mathrm{ns}\). This timescale is fully compatible with conventional electro- and acousto-optic modulators, allowing the amplitudes and phases of the optical tones in each interaction zone to be reconfigured between successive round trips without difficulty. A magnetic field of order a few millitesla (\(B\simeq 2~\mathrm{mT}\)) is sufficient to realize such an orbit, well within standard SEM capabilities. Alternatively, the electrons may be guided with the help of Paul trap-like structures~\cite{Hoffrogge2011,Zimmermann2021}, where even electron resonators have recently been demonstrated~\cite{Seidling2024} .In practice, longitudinal dispersion of the electron wavepacket over many revolutions must be taken into account; however, established techniques such as magnetic compression, energy chirping, or tailored dispersion compensation can be employed to mitigate wavepacket spreading and preserve coherence. After completing the programmed sequence of interactions, the quantum dynamics are encoded in the electron energy spectrum, which can be read out using an electron energy spectrometer and a microchannel plate (MCP) or a similar single-electron-sensitive detector.

\begin{figure}[ht]
\centering
\includegraphics[width=0.95\linewidth]{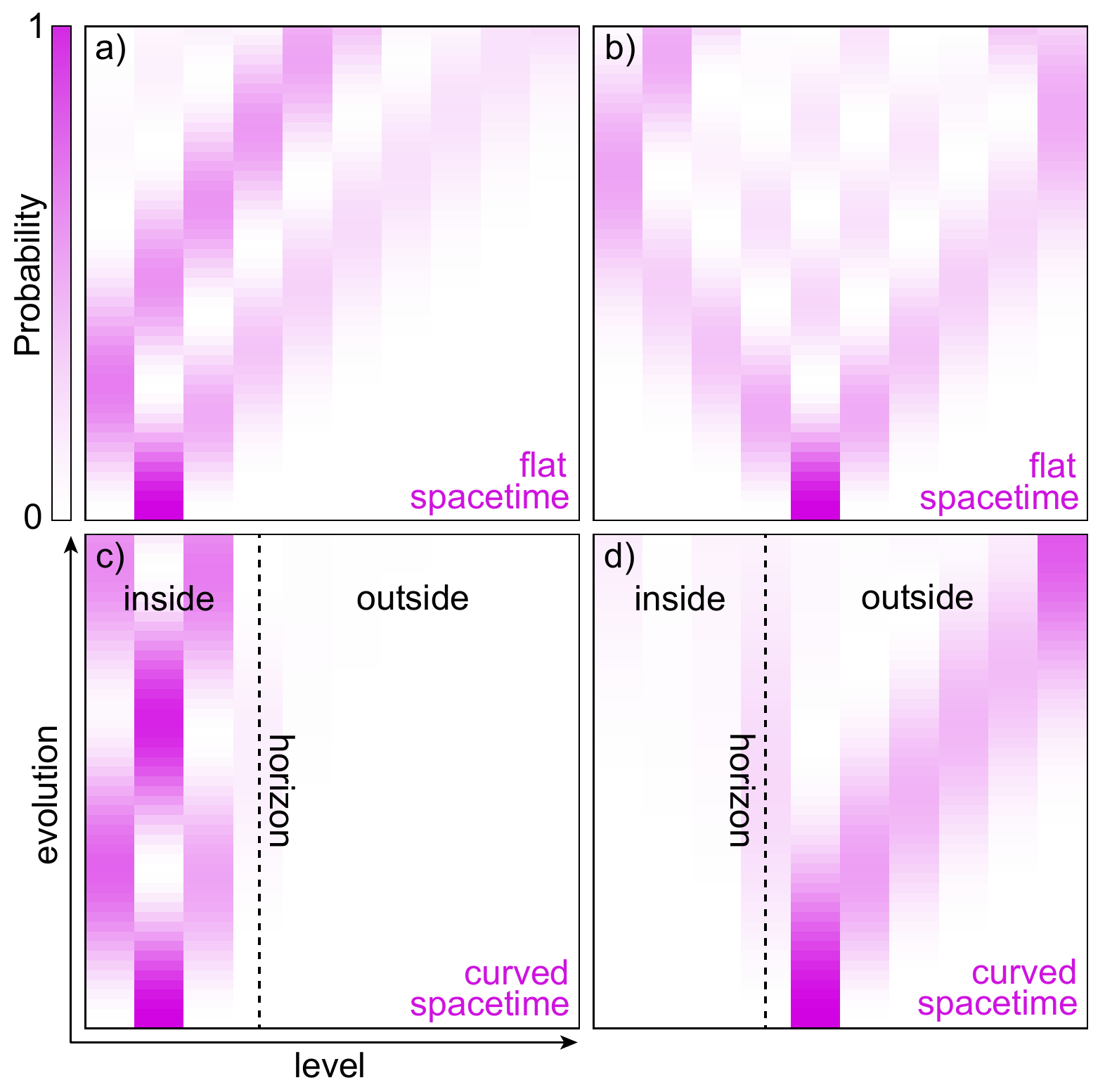}
\caption{\label{fig:BH} Quantum walks in flat and curved spacetime.
{a,b)} Quantum walk dynamics on an effective flat spacetime with uniform ladder couplings.
{c,d)} Quantum walk dynamics on a curved spacetime corresponding to a one-dimensional analogue black hole, with a horizon located at ladder index \(m=3\).
The horizon induces asymmetric propagation: excitations initialized inside the horizon remain trapped ({c}, \(m_0=1\)), while those initialized outside propagate away ({d}, \(m_0=4\)).}
\end{figure}

As shown in Fig.~\ref{fig:QC}b, the recoil-induced energy ladder forms a high-dimensional qudit with full and universal control over its nearest-neighbor couplings. By programming the amplitudes, phases, and spatial arrangement of the optical drives, one can engineer a wide class of effective ladder Hamiltonians within a single propagating particle. This capability goes beyond static tight-binding models and allows the simulation of position-dependent and nonuniform coupling landscapes that emulate effective spacetime geometries (see Fig.~\ref{fig:BH} and the following discussion).

Figure~\ref{fig:QC}c highlights the use of the ladder qudit for scalable quantum information processing. A central advantage of the recoiled-electron platform is that multiple logical qubits can be encoded within a single physical system by selecting appropriate subsets of ladder states. In general, an effective register of \(N\) qubits can be embedded in a ladder qudit using \(2^N\) energy levels. This encoding allows complex multiqubit operations to be performed within a single, highly coherent quantum object, avoiding the overhead associated with coupling many separate qubits.

In the example shown, two logical qubits are encoded using four consecutive ladder states, \(\ket{m}\)–\(\ket{m+3}\). The figure reports the performance of both qudit-level operations and encoded-qubit gates as a function of the interaction length per zone. Elementary operations, such as a nearest-neighbor swap \(\mathrm{SWAP}_{\ket{m}\rightarrow\ket{m+1}}\), can be implemented using a single interaction zone. This swap operation is native to the recoiled-electron ladder, as it directly corresponds to the fundamental nearest-neighbor coupling. As such, it provides a natural operation within the ladder and serves as a key building block for more complex gates. More elaborate transformations---such as the next-nearest-neighbor swap \(\mathrm{SWAP}_{\ket{m}\rightarrow\ket{m+2}}\)---are constructed from sequences of nearest-neighbor swaps and therefore require multiple interaction zones to refocus intermediate levels and suppress leakage. These swap-based operations form the backbone of universal control in the ladder qudit, enabling the compilation of arbitrary multiqubit and multilevel gates (see Supplementary Information \Cref{sec:Supp_gates}).

A key ingredient for universal multiqubit control is the availability of high-fidelity entangling gates. As a representative example, we implement a controlled-phase gate, \(\mathrm{CZ}\), using two complementary approaches. In the first, a single \(2\pi\) interaction involving an auxiliary ladder level \(\ket{m+4}\) produces the desired conditional phase (orange curve in Fig.~\ref{fig:QC}c). In the second, higher-fidelity \(\mathrm{CZ}\) operations are realized using composite pulse sequences optimized over multiple interaction zones (Fig.~\ref{fig:QC}c, stars \(\mathrm{CZ}^*\)). Specifically, we employ optimal control over several interaction zones, while restricting the maximum length \(L_{\rm max}\) of a single interaction zone. We optimize over seven interaction zones with \(L_{\rm max}=10^{-4}\,\mathrm{m}\) (left star), and over nine interaction zones with \(L_{\rm max}=10^{-3}\,\mathrm{m}\) (right star). These multi-zone sequences suppress residual phase errors and leakage, achieving substantially higher fidelities than simple single-zone implementations.

The dominant error mechanism in these gates arises from population leakage outside the encoded subspace. In the recoiled-electron ladder, this leakage decreases quadratically with the interaction length, leading to an overall error scaling proportional to \(1/L^2\). As a result, increasing the interaction length or distributing the operation across multiple zones enables systematic improvement of gate fidelity. Together, these results demonstrate that the ladder qudit supports high-fidelity multiqubit operations with favorable scaling, making recoiled free electrons a promising platform for compact and scalable quantum information processing. 

Beyond gate-based operations, the same programmable recoil ladder can also be used as an analogue quantum simulator. To demonstrate this capability, we present an example in which the ladder dynamics emulate wave propagation in a one-dimensional curved spacetime, realizing an analogue black-hole Hamiltonian with a synthetic horizon.

In this setting, the ladder index \(m\) is interpreted as a discrete synthetic spatial coordinate, while
the propagation coordinate \(z\) plays the role of time for the simulated dynamics.
By tailoring the optical-tone amplitudes and phases, one can engineer position-
dependent hopping amplitudes along this synthetic dimension, thereby realizing
effective lattice Hamiltonians that emulate wave propagation in curved spacetime.

Figure~\ref{fig:BH} illustrates this idea for a one-dimensional analogue black-hole
Hamiltonian associated with a Painlevé--Gullstrand metric containing a smooth
Rindler horizon \cite{faccio2013analogue, shi2023quantum}. In the corresponding
tight-binding description, wave packets propagate along the synthetic coordinate
\(m\). The relevant group velocity is therefore not the physical longitudinal
velocity of the electron, but the effective velocity \(dm/dz\) of a wave packet in
the recoil ladder, determined by the engineered band structure and coupling profile.

The analogue horizon occurs at the synthetic position where the lab-frame group
velocity of an outward-propagating mode vanishes. Equivalently, this is the point
where the engineered effective flow cancels the intrinsic propagation velocity of
the lattice excitation, so that modes on one side of the horizon cannot propagate
back across it. Experimentally, this horizon is implemented by programming a
spatially varying ladder-coupling profile, for example through a rapid variation
or sign change of the effective hopping amplitudes. In the simulations shown, the
interaction length is fixed to \(L = 10^{-3}\,\mathrm{m}\), and the effective
evolution is generated by gradually increasing the magnitude of the applied optical
couplings \(|J|\).



To demonstrate the resulting horizon physics, we prepare an initial excitation localized either inside or outside the effective horizon. When initialized inside the horizon, the excitation remains trapped and cannot propagate outward, reflecting the unidirectional flow imposed by the curved spacetime geometry. In contrast, an excitation prepared outside the horizon propagates away from the black hole, demonstrating the asymmetric transport characteristic of horizon physics. This behavior directly mirrors the causal structure of black-hole spacetimes, where information inside the horizon is prevented from escaping.

The simulation is implemented using simultaneously applied optical tones, with selectivity arising from momentum mismatch in the high-recoil regime. Under these conditions, the dynamics reduce to an effective autonomous Hamiltonian \(H_{\mathrm{eff}}\) that faithfully reproduces the desired curved-spacetime lattice model. Importantly, because the optical tones can be controlled individually and modulated on ultrafast timescales, this platform naturally enables the study of time-dependent curved-spacetime geometries, including dynamical horizons and evolving metric profiles, realized by selectively addressing different ladder levels and modulating their couplings to induce spatially varying dispersion relations with horizon-like turning points. Such rapid control provides a direct analogue of cosmological particle production and dynamical Casimir--type effects \cite{jaskula2012acoustic}, in which time-dependent backgrounds generate excitations from the vacuum. In this respect, the recoiled free-electron platform offers a significant advantage over superconducting-circuit implementations \cite{shi2023quantum}, where the achievable timescales for dynamically varying the effective metric are typically much slower. Beyond single-particle transport, the same framework supports investigations of mode conversion, horizon-induced correlations, and Hawking-like processes, providing a scalable route to analogue-gravity simulations with recoiled free electrons. Full details of the black-hole Hamiltonian, coupling profiles, and numerical simulation protocol are provided in the Supplementary Information \Cref{sec:Supp_black_holes}.

\section{Quantum state generation with recoiled free electrons}\label{sec:QO}

\begin{figure*}[ht]
\centering
\includegraphics[width=0.75\linewidth]{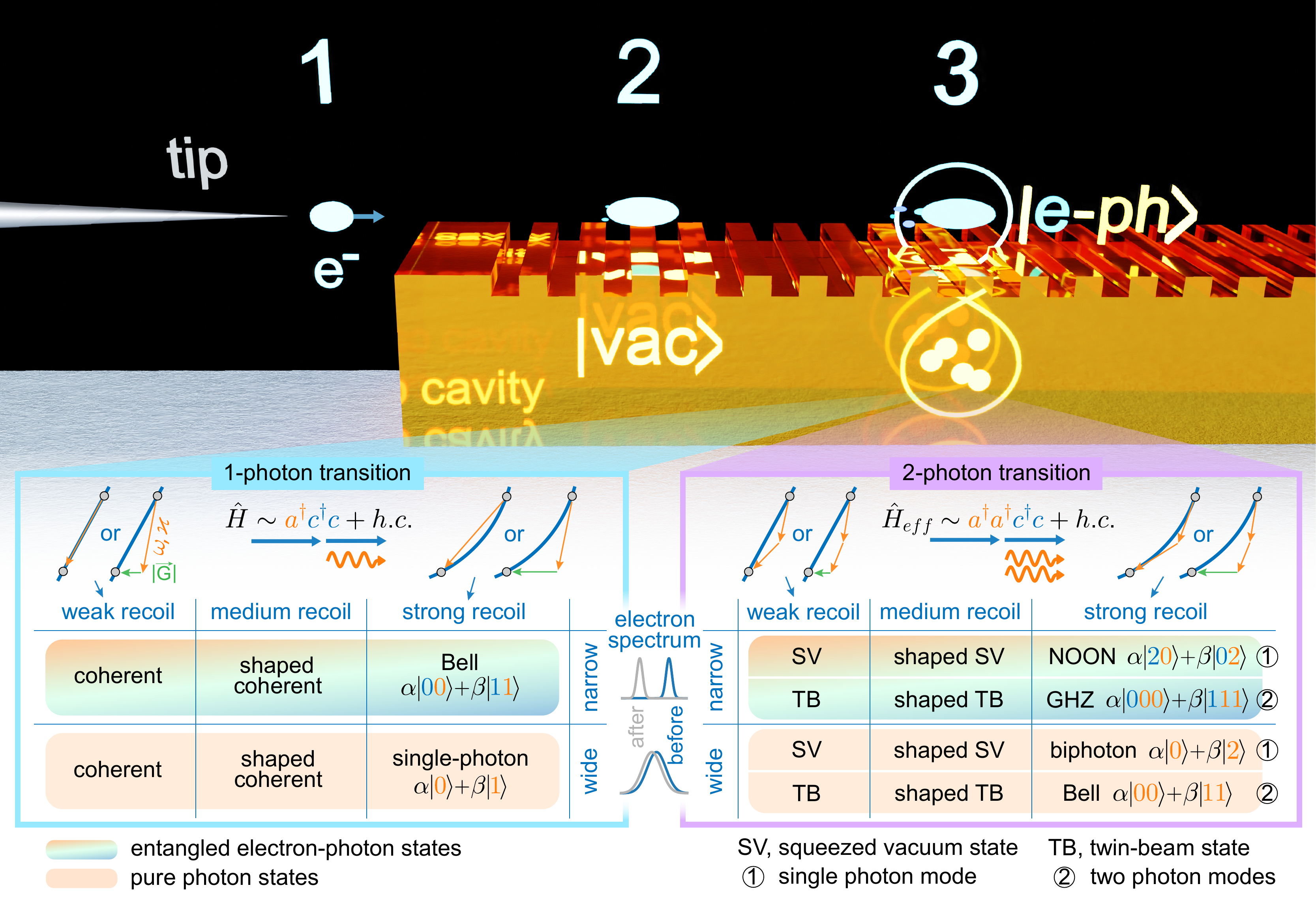}
\caption{\label{fig:abstract} Schematic summary of electron--photon state generation.
{(1)} Free-electron emission (e.g., from a metallic nanotip).
{(2)} Interaction with the evanescent vacuum field of an optical cavity, \(\ket{\mathrm{vac}}\).
{(3)} Formation of an entangled electron--photon state, \(\ket{\mathrm{e\!-\!ph}}\).
Depending on the interaction order (one- or two-photon), recoil strength, and electron spectral width, a variety of quantum states can be generated. The table summarizes the role of electron dispersion (linear: weak recoil; nonlinear: strong recoil) and electron spectral width (overlapping versus distinguishable spectra before and after interaction).
}
\end{figure*}

We now turn to a complementary regime of the same electron--cavity system, in which the cavity is initially prepared in the vacuum state or in low-lying photon Fock states, in the absence of an external coherent drive (Fig.~\ref{fig:abstract}). In this case, the interaction between a flying electron and the quantized near field of an empty cavity leads to the generation of entangled electron--photon states. As the electron traverses the cavity, it may emit photons into the cavity mode or absorb photons from it, provided that the phase-matching condition between the electron momentum change and the photon momentum is satisfied (see additionally Fig.~\ref{fig:dispersion}a). The nature of the resulting quantum states depends sensitively on the electron dispersion. For relativistic electrons with nearly linear dispersion, recoil is weak and photon emission leads to approximately Poissonian statistics \cite{kfir_entanglements_2019}. In contrast, for slow electrons with parabolic dispersion, quantum recoil becomes significant and suppresses consecutive emission of photons with the same frequency due to accumulated phase mismatch \cite{huang_quantum_2023}. As a result, subsequent photon emission requires different frequencies, giving rise to nonclassical photon statistics and structured hybrid light--matter states. In the following sections, we analyze this vacuum-coupling regime in detail and show how quantum recoil enables the deterministic generation of a broad class of nonclassical electron--photon and purely photonic states.

To quantitatively describe these processes, we consider a fully quantized interaction Hamiltonian of the form
\({H} \sim g_{\rm Qu}\,\hat{a}^{\dagger}\hat{c}^{\dagger}\hat{c} + \mathrm{h.c.}\),
where \(\hat{c}^{\dagger}\) and \(\hat{a}^{\dagger}\) are the electron and photon creation operators, respectively, and \(g_{\rm Qu}\) is the vacuum electron--photon coupling strength. If the cavity is initially in the vacuum state \(\ket{0}\) and the electron has initial energy \(E_0=\hbar\Omega_0\), the joint state can be expanded as
\(\ket{\psi}=\sum_{n\ge 0} C_{-n}\ket{-n,n}\),
where \(\ket{-n,n}\equiv \ket{E_0-n\hbar\omega}\otimes\ket{n}\) denotes an electron that has lost \(n\) photons of energy \(\hbar\omega\) (so \(\Delta\Omega_0=n \omega\)) while the cavity occupies the \(n\)-photon Fock state \(\ket{n}\). We refer to the state \(\ket{-n,n}\) as the \(n^{\mathrm{th}}\) level of the electron--photon system. In the following, we solve this problem numerically and, in appropriate limits, obtain exact analytical solutions after autonomization, allowing for a direct comparison between the two approaches.

We introduce the phase-matching (PM) width \(\sigma\) as the recoil parameter, which characterizes how many photons an electron can emit before recoil-induced phase mismatch becomes significant:
\begin{equation}
\sigma \;\approx\;
155\,\frac{\left[E_{\rm kin}\,(\mathrm{keV})\right]^{3/2}}
{\left[E_{\rm ph}\,(\mathrm{eV})\right]^2\,L\,(\mu\mathrm{m})},
\label{eq:sigma}
\end{equation}
where \(E_{\rm ph}=\hbar\omega\) is the photon energy, \(L\) is the interaction length, and \(E_{\rm kin}\) denotes the kinetic energy of the electron (see Supplementary Information \Cref{sec:Supp_elphot_states} for details).

The effective number of participating levels \(N_{\mathrm{eff}}\) in the electron--photon system can be estimated from the number of emission (\(\Delta\Omega/\omega<0\)) or absorption (\(\Delta\Omega/\omega>0\)) sidebands, together with the zero-loss peak at \(\Delta\Omega/\omega=0\), as
\begin{equation}
N_{\mathrm{eff}}=
\begin{cases}
\sigma+1, & \sigma\ge 1,\\[4pt]
2, & \sigma<1.
\end{cases}
\label{eq:neff}
\end{equation}
Figure~\ref{fig:1photon}a shows the PM width \(\sigma\) as a function of electron kinetic energy \(E_{\rm kin}\) and photon vacuum wavelength \(\lambda\). For small electron energies or short photon wavelengths, recoil rapidly drives the system away from phase matching, resulting in an effective two-level system for \(\sigma<1\). With increasing electron energy or longer wavelengths, more ladder levels become accessible, eventually approaching the infinite-ladder limit.

To characterize the photon statistics of the electron--photon state, we consider the normalized second-order correlation function at zero delay,
\[
g^{(2)}(0)=
\frac{\langle\hat{a}^\dagger\hat{a}^\dagger\hat{a}\hat{a}\rangle}
{\langle\hat{a}^\dagger\hat{a}\rangle^2},
\]
which exhibits antibunching for \(g^{(2)}<1\) (sub-Poissonian statistics), bunching for \(g^{(2)}>1\) (super-Poissonian statistics), and Poissonian statistics for \(g^{(2)}=1\). Values \(g^{(2)}>3\) correspond to superbunched emission.

\subsection{Single-photon processes}

\begin{figure*}
\centering
\includegraphics[width=1\linewidth]{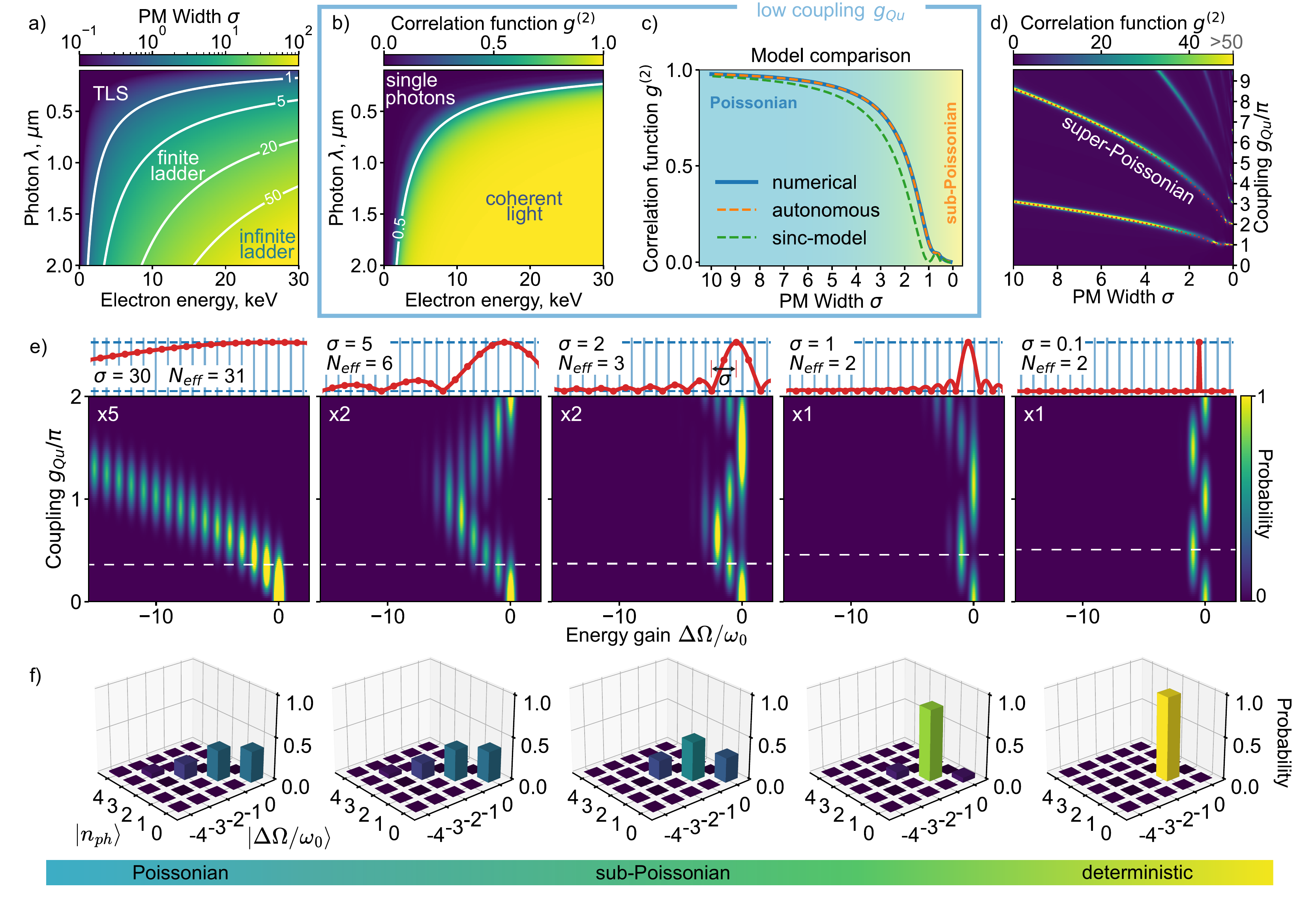}
\caption{\label{fig:1photon} Single-photon processes.
{a)} Phase-matching (PM) width \(\sigma\) as a function of photon wavelength and electron energy, illustrating the transition from an infinite ladder to a finite ladder and ultimately to an effective two-level system at lower electron energies.
{b)} Second-order correlation function \(g^{(2)}\) as a function of photon wavelength and electron energy, showing the crossover from coherent to single-photon generation.
{c)} Comparison of theoretical models: \(g^{(2)}\) versus PM width \(\sigma\), demonstrating the transition from Poissonian to sub-Poissonian statistics.
{d)} \(g^{(2)}\) as a function of coupling strength \(g_{\rm Qu}\) and PM width \(\sigma\), revealing super-Poissonian statistics associated with multi-level Rabi oscillations.
{e)} Electron energy spectra for varying \(g_{\rm Qu}\) and \(\sigma\), showing the transition from Poissonian to deterministic single-photon generation (insets: relative level couplings).
{f)} Joint electron--photon probability distributions corresponding to the maximal single-photon generation points marked in (e).
The interaction length is \(L=500~\mu\mathrm{m}\) for panels (a,b), and \(\omega_0 \equiv \omega\).
}
\end{figure*}

We first consider single-photon transitions in the low-coupling regime (\(g_{\rm Qu} \lesssim 0.5\)). In the absence of recoil (\(\sigma \gg 1\)), the electron--photon state takes the form of a weak coherent state,
\begin{equation}
\ket{\psi} = \sum_{n=0}^{\infty} e^{-|g_{\rm Qu}|^2/2}\,
\frac{g_{\rm Qu}^n}{\sqrt{n!}}\ket{-n,n}
\equiv \ket{\alpha^{\mathrm{e\text{-}ph}}},
\end{equation}
as previously reported \cite{kfir_entanglements_2019}. As the recoil parameter \(\sigma\) decreases, recoil suppresses population of higher-energy ladder states, producing a sharp cutoff in the distribution for \(n \gtrsim \sigma\). In the extreme recoil-dominated limit \(\sigma \lesssim 1\), the system effectively reduces to a two-level subspace,
\(\ket{\psi} \approx c_0\ket{0,0} + c_1\ket{-1,1}\),
with \(c_1 \approx g_{\rm Qu}\).

The photon subsystem of this state can function as a single-photon source. As \(\sigma\) decreases, we observe a transition from Poissonian to sub-Poissonian photon statistics with clear antibunching (\(g^{(2)}<1\)), as shown in Fig.~\ref{fig:1photon}c. For experimentally realistic values of \(E_{\rm kin}\), \(E_{\rm ph}\), and \(L\), recoil-induced antibunching can be substantial (Fig.~\ref{fig:1photon}b). For example, at \(E_{\rm ph}=2.33~\mathrm{eV}\) (vacuum wavelength \(532~\mathrm{nm}\)), with \(5~\mathrm{keV}\) electrons and an interaction length of \(0.4~\mathrm{mm}\), we obtain \(\sigma \simeq 0.83\) and \(g^{(2)} \simeq 0.05\), comparable to state-of-the-art quantum-dot single-photon sources \cite{senellart_high-performance_2017}. Figures~\ref{fig:1photon}b,c demonstrate that in the low-coupling regime the recoil parameter \(\sigma\) alone determines the photon statistics.

In the high-coupling regime (\(g_{\rm Qu} \gtrsim 0.5\)), the electron--photon dynamics become richer and depend on both \(\sigma\) and \(g_{\rm Qu}\). As \(\sigma\) decreases, the system evolves from the coherent state \(\ket{\alpha^{\mathrm{e\text{-}ph}}}\) toward a new class of recoil-shaped states, which we denote as \(\ket{\alpha^{\mathrm{e\text{-}ph}}_{\mathrm{shaped}}}\) (Fig.~\ref{fig:1photon}e,f). These states exhibit oscillatory behavior as a function of \(g_{\rm Qu}\): population accumulates near the recoil cutoff \(n \approx \sigma\), reflects from this boundary, and returns toward the zero-loss level. Increasing \(g_{\rm Qu}\) leads to repeated oscillations with growing amplitude (see Extended Data Fig.~\ref{fig:superbunching}).

In the strong recoil limit \(\sigma \lesssim 1\), the shaped coherent state reduces to a Bell-like state,
\(\ket{\psi} = \alpha\ket{0,0} + \beta\ket{-1,1}\),
with \(\alpha=\cos(g_{\rm Qu})\) and \(\beta=\sin(g_{\rm Qu})\), exhibiting clear Rabi oscillations (Fig.~\ref{fig:1photon}e,f and Extended Data Fig.~\ref{fig:GHZ}). By tuning \(g_{\rm Qu}\), one can deterministically generate the single-photon state \(\ket{\psi}=\ket{-1,1}\) or the maximally entangled electron--photon Bell state
\(\ket{\Phi^+}=(\ket{00}+\ket{11})/\sqrt{2}\),
where the electron state \(\ket{-1}\) is relabeled as \(\ket{1}\) for clarity.

Depending on the values of \(\sigma\) and \(g_{\rm Qu}\), the photon subsystem of the shaped electron--photon state \(\ket{\alpha^{\mathrm{e\text{-}ph}}_{\mathrm{shaped}}}\) can exhibit Poissonian, sub-Poissonian, or super-Poissonian statistics (Fig.~\ref{fig:1photon}d). Pronounced peaks in the \(g^{(2)}\) map coincide with revivals of the zero-loss population and correspond to the generation of superbunched \(k\)-photon states with low mean photon number \(\langle \hat{N} \rangle \ll 1\) (see Extended Data Fig.~\ref{fig:superbunching}).

Finally, we note that a free electron can act as a deterministic single-photon absorber when the cavity is initially prepared in the one-photon Fock state \(\ket{1}\) (Extended Data Fig.~\ref{fig:absorption}). Moreover, the post-interaction electron energy spectrum provides a sensitive probe of the cavity state, allowing discrimination between Fock, coherent, and thermal photon statistics \cite{dahan_imprinting_2021-1, giulio_probing_2019}.

\subsection{Two-photon processes}

\begin{figure*}
\centering
\includegraphics[width=1\linewidth]{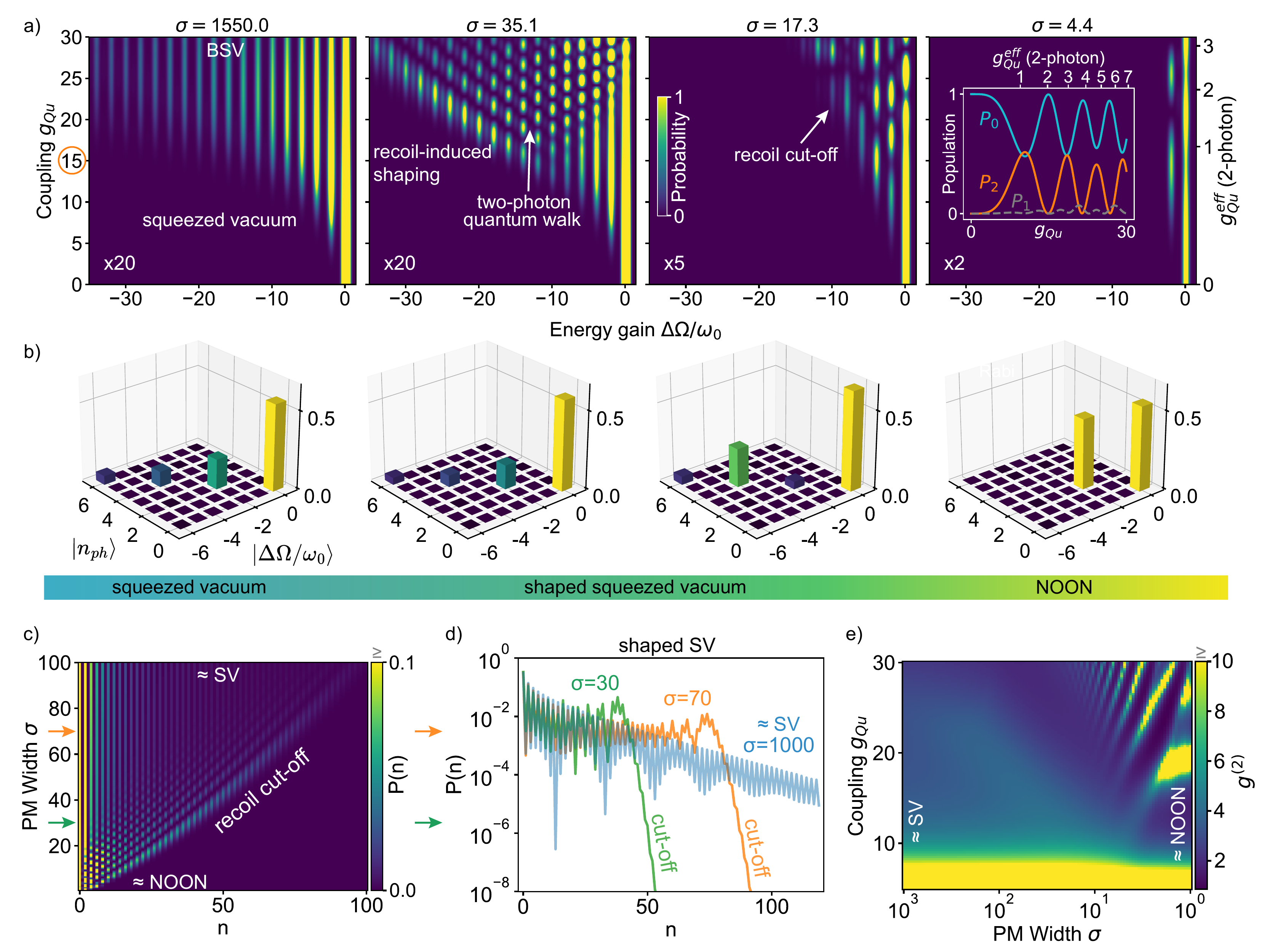}
\caption{\label{fig:bsv} Two-photon processes.
\textbf{a)} Electron energy spectra as a function of coupling strength \(g_{\rm Qu}\) for different phase-matching (PM) widths \(\sigma\), showing the transition from electron--photon squeezed vacuum (SV) to recoil-shaped SV and NOON-like states (the two-photon transition is phase-matched as in Fig.~\ref{fig:dispersion}f). The inset shows two-photon Rabi oscillations for reduced one-photon mismatch \(\varphi_{-1}L\).
\textbf{b)} Joint electron--photon probability distributions corresponding to the states in (a) for \(g_{\rm Qu}=15\).
\textbf{c,d)} Photon-number distributions \(P(n)\) of the electron--photon state for different \(\sigma\) at fixed \(g_{\rm Qu}=22\).
\textbf{e)} Second-order correlation function \(g^{(2)}\) as a function of \(g_{\rm Qu}\) and \(\sigma\).
}
\end{figure*}

We now investigate two-photon near-field Cherenkov radiation processes (Fig.~\ref{fig:dispersion}f). In this regime, single-photon emission or absorption is phase-mismatched, while the simultaneous emission or absorption of two photons satisfies the phase-matching condition and therefore dominates the dynamics.

Such two-photon processes can be approximately described by an effective Hamiltonian of the form
\[
{H}_{\mathrm{eff}} \sim g_{\rm Qu}^{\mathrm{(2)}}\,\hat{a}^\dagger\hat{a}^\dagger\hat{c}^\dagger\hat{c} + \mathrm{h.c.},
\]
where the effective two-photon coupling strength is given by
\[
g_{\rm Qu}^{\mathrm{(2)}} \equiv \frac{g_{\rm Qu}^2}{-\varphi_{-1}L}.
\]
Here \(\varphi_{-1}L=-\Delta k_{-1}L=-(k(\Omega_0)-k(\Omega_{-1})-\varkappa)L\) denotes the one-photon transition mismatch, and \(g_{\rm Qu}\) is the single-photon vacuum coupling strength.

We first consider the non-recoil regime \(\sigma\gg1\). Because the two-photon transition is phase-matched, the resulting electron--photon state contains predominantly even photon numbers and even electron energy-loss states,
\[
\ket{\psi}\approx c_0\ket{0,0}+c_1\ket{-2,2}+c_2\ket{-4,4}+\cdots.
\]
We refer to this state as an electron--photon squeezed vacuum, \(\ket{\mathrm{SV}^{\mathrm{e\text{-}ph}}}\) (see Supplementary Information \Cref{sec:Supp_elphot_states} for details). By increasing the effective coupling \(g_{\rm Qu}^{\mathrm{(2)}}\), the system evolves from the weak photon-pair regime \(|c_1|\ll|c_0|\), with \(c_1\simeq g_{\rm Qu}^{\mathrm{(2)}}\), to a bright squeezed-vacuum state (Fig.~\ref{fig:bsv}a,b, left panels).

As the recoil parameter \(\sigma\) decreases, the electron--photon state undergoes recoil-induced shaping. We denote the resulting states as recoil-shaped squeezed vacuum,
\(\ket{\mathrm{SV}^{\mathrm{e\text{-}ph}}_{\mathrm{shaped}}}\)
(Fig.~\ref{fig:bsv}a,b, central panels). Recoil suppresses population of high-\(n\) components, producing a sharp cutoff in the photon-number distribution. The position of this cutoff depends on both the coupling strength \(g_{\rm Qu}^{\mathrm{(2)}}\) (Fig.~\ref{fig:bsv}a) and the recoil parameter \(\sigma\) (Fig.~\ref{fig:bsv}c). A further consequence of recoil is the emergence of a two-photon quantum walk, visible in Fig.~\ref{fig:bsv}a, arising from interference of probability amplitudes reflected at the cutoff boundary.

In the strong-recoil limit \(\sigma\lesssim1\), the shaped squeezed-vacuum state reduces to a NOON-like superposition,
\[
\ket{\psi}\approx \alpha\ket{0,0}+\beta\ket{-2,2},
\]
with \(\alpha=\cos(g_{\rm Qu}^{\mathrm{(2)}})\) and \(\beta=\sin(g_{\rm Qu}^{\mathrm{(2)}})\), exhibiting two-photon Rabi oscillations (Fig.~\ref{fig:bsv}a,b, right panels and Extended Data Fig.~\ref{fig:GHZ}). In this regime, the dynamics are well described by an effective three-level system involving the coupled states \(\ket{0,0}\) and \(\ket{-2,2}\), with the intermediate state \(\ket{-1,1}\) remaining off-resonant, analogous to a Raman process \cite{scully_quantum_1997}. At \(g_{\rm Qu}^{\mathrm{(2)}}\simeq1\), a maximally entangled electron--photon NOON state,
\(\ket{\mathrm{NOON}}=(\ket{20}+\ket{02})/\sqrt{2}\),
can be generated.

Control over \(\sigma\) and \(g_{\rm Qu}\) (or equivalently \(g_{\rm Qu}^{\mathrm{(2)}}\)) enables systematic engineering of photon statistics. As shown in Fig.~\ref{fig:bsv}c--e, the recoil introduces a cutoff that suppresses the heavy tails characteristic of squeezed-vacuum states. In the absence of recoil (\(\sigma\gg1\)), the photon statistics satisfy \(g^{(2)}=3+1/\langle\hat{N}\rangle\), as expected for squeezed vacuum \cite{klyshko_photons_2017}. As \(\sigma\) decreases, \(g^{(2)}\) develops pronounced oscillatory features, allowing the degree of photon bunching to be tuned sensitively via system parameters. At the same time, both \(\ket{\mathrm{SV}^{\mathrm{e\text{-}ph}}}\) and \(\ket{\mathrm{SV}^{\mathrm{e\text{-}ph}}_{\mathrm{shaped}}}\) exhibit strong correlations between electron and photon subsystems, enabling photon-number resolution by measuring the electron spectrum and facilitating the preparation of non-Gaussian states of light \cite{virally_enhanced_2021}.

If the cavity supports two modes that are jointly phase-matched to the two-photon transition (one photon emitted into each mode), the same mechanism produces an electron--photon twin-beam state, or two-mode squeezed vacuum, \(\ket{\mathrm{Twin}^{\mathrm{e\text{-}ph}}}\) (see Extended Data Fig.~\ref{fig:twin}). In the strong-recoil limit \(\sigma\lesssim1\), this state evolves into a GHZ-like superposition \(\alpha\ket{000}+\beta\ket{111}\) (Extended Data Fig.~\ref{fig:GHZ} and similar to before: the first entry represents the electron, the following two refer to the two optical modes), with the maximally entangled GHZ state
\(\ket{\mathrm{GHZ}}=(\ket{000}+\ket{111})/\sqrt{2}\)
obtained at \(g_{\rm Qu}^{\mathrm{(2)}}\simeq1\). Deterministic generation of degenerate and nondegenerate photon pairs, \(\ket{\psi}=\ket{-2,2,0}\) and \(\ket{\psi}=\ket{-2,1,1}\), respectively, is also possible for low \(\sigma\), small one-photon mismatch, and \(g_{\rm Qu}^{\mathrm{(2)}}\simeq\pi/2\) (Extended Data Fig.~\ref{fig:3lvl}d,e).

Finally, our framework naturally extends to one- and two-photon PINEM interactions with strong classical electromagnetic fields and recoiled electrons (Extended Data Fig.~\ref{fig:pinem}). Because PINEM can modulate the electron wavefunction and generate attosecond electron bunch trains \cite{feist_quantum_2015, kozak_inelastic_2018, schonenberger_generation_2019, black_net_2019}, incorporating recoil and two-photon phase matching provides additional control knobs for electron wavefunction shaping and temporal compression. Two-photon PINEM can also lead to photon-state squeezing, analogous to ponderomotive interactions \cite{giulio_optical-cavity_2022}. In existing PINEM experiments \cite{kfir_controlling_2020, henke_integrated_2021}, the recoil parameter \(\sigma\) was typically much larger than the observed number of sidebands, suppressing recoil effects. PINEM in the SEM regime \cite{shiloh_quantum-coherent_2022} therefore represents a promising route for the first experimental observation of recoil-induced shaping in PINEM spectra.

\section{Outlook}

In this work, we have investigated the role of quantum electron recoil in free-electron--photon interactions and shown that recoil fundamentally enriches the accessible quantum dynamics. We developed a first-principles quantum electrodynamics framework that treats recoil exactly and used it to reveal a universal recoil-induced ladder structure governed by a single dimensionless parameter \(\sigma\), which depends only on the electron energy, photon energy, and interaction length.

This ladder structure establishes recoiled free electrons as a versatile platform for quantum simulation and Hamiltonian engineering. As demonstrated in the first part of this work, programmable couplings between recoil states enable the realization of effective lattice models within a single propagating particle, including curved-spacetime and analogue black-hole Hamiltonians. Looking forward, this platform opens opportunities to explore more complex dynamical spacetimes, non-equilibrium lattice models, and interacting quantum walks.

Beyond quantum simulation, the same recoil-enabled dynamics provide a powerful resource for quantum state generation. Using our framework, we analyzed the deterministic generation of a broad class of nonclassical electron--photon and purely photonic states. In particular, we presented protocols for creating electron--photon and pure-photon Bell, NOON, GHZ, squeezed-vacuum, and twin-beam states, as well as more general hybrid light--matter states enabled by recoil. Our results demonstrate deterministic single-photon generation and absorption, along with deterministic degenerate and nondegenerate photon-pair creation. Together, these capabilities position recoil-based free-electron platforms as a unified setting in which quantum simulation and hybrid light--matter state engineering naturally coexist.

Our work also raises several open questions for future investigation. These include: (i) how do squeezing processes manifest not only in the phoronic subsystem, but in the electron wavefunction itself, for example through energy--momentum correlations or quadrature-like observables; (ii) how does recoil-induced squeezing modify the dynamics of cavities initially prepared in coherent or thermal states; and (iii) how do recoil and multiphoton processes influence the temporal and spectral properties of attosecond electron bunches in PINEM and related schemes? We anticipate that the framework developed here will provide a natural starting point for addressing these questions and for extending free-electron platforms toward quantum simulation, quantum-enhanced sensing, and compact quantum information processing architectures. Finally, because all key ingredients—electron sources, optical cavities, magnetic control, and energy-resolved detection—are naturally compatible with modern scanning electron microscopes, our approach offers a realistic and scalable path toward experimental implementations of quantum simulation and hybrid light--matter control in free-electron systems.

\section*{Supplementary information}
Supplementary files are attached.

\section*{Data availability}
The data supporting the findings of this study are available from the
corresponding authors upon reasonable request.

\section*{Code availability}
The code supporting the findings of this study is available from the
corresponding authors upon reasonable request.

\section*{Acknowledgments}
This research was supported by the European Research Council Advanced Grant
AccelOnChip, the Gordon and Betty Moore Foundation Grant No.~11473, and the
Deutsche Forschungsgemeinschaft Project-ID~429529648: TRR~306 QuCoLiMa.

\section*{Author contributions}
M.S. and P.H. initiated the research. M.S. and A.R. derived the theory.
M.S. performed the numerical calculations and wrote the first draft of the
manuscript. The project was supervised by T.C., R.S., and P.H. All authors
discussed the results and contributed to the writing of the manuscript.

\section*{Competing interests}
The authors declare no competing interests.

\bibliography{sample}


%% file: supplement.tex
\section*{System description}
\section{Microscopic picture of recoil}\label{sec:Supp_recoil}
\begin{figure*}[ht]
\centering
\includegraphics[width=1\linewidth]{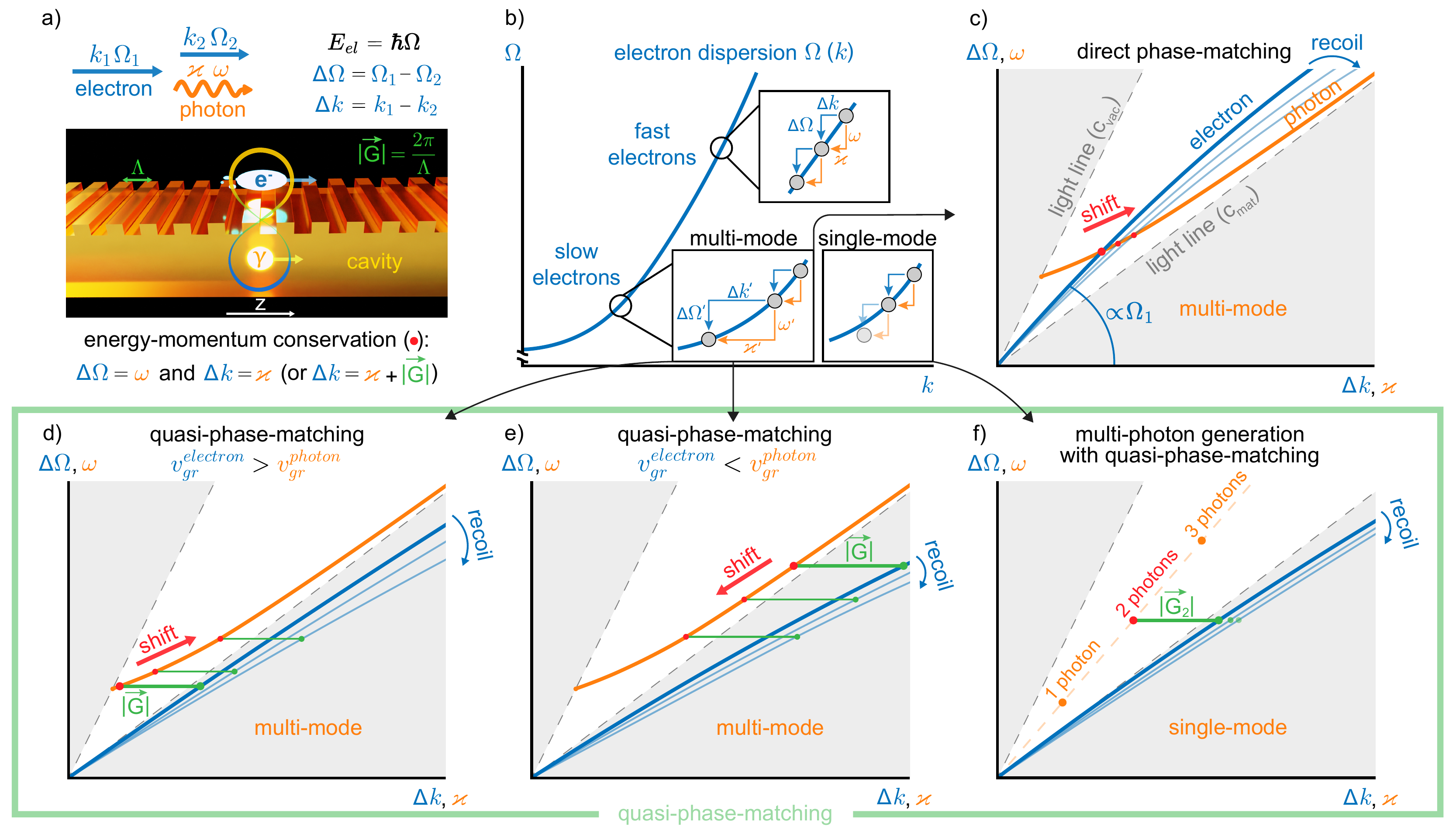}
\caption{\label{fig:dispersion} Dispersion curves and phase-matching condition. a) Visualization and diagram of the process: initial electron with energy $\hbar \Omega_1$ and momentum $\hbar k_1$ annihilates and generates photons with angular frequency $\omega$ and momentum $\hbar \varkappa$ and a final electron with energy $\hbar \Omega_2$ and momentum $\hbar k_2$. Energy-momentum conservation leads to the conditions $\Omega_1 - \Omega_2 \equiv \Delta \Omega = \omega$ and $k_1 - k_2 \equiv \Delta k = \varkappa$. b) Electron dispersion curve $\Omega(k)$ for fast electrons (upper inset) and slow electrons (lower inset). c-f) Schematic dependence of the electron energy loss $\hbar \Delta \Omega$ on the electron momentum loss $\hbar \Delta k$ for the initial electron energy $\hbar \Omega_1$ (bold blue curve) and after the consecutive emission of photons (thin blue curves), and the schematic dispersion curve $\omega(\varkappa)$ of a cavity photon (orange curve). d) Recoil with quasi-phase-matching (QPM) and $v^{electron}_{gr}>v^{photon}_{gr}$ results in a blue shift of the generated photon. $\vec{G}$ (green) is a diffraction grating vector. e) Recoil with QPM and $v^{electron}_{gr}<v^{photon}_{gr}$ results in a red shift of the generated photon. (c), (d) and (e) assume a multi-mode cavity. f) QPM allows the generation of k-photon states inside the single-mode cavity.  Here, for example, two-photon (k=2) generation is phase-matched.}
\end{figure*}

We consider the interaction of free electrons with the evanescent, quantized electric field of a cavity (Fig.~\ref{fig:dispersion}a). Free electrons, generated in an SEM or TEM, on-chip or with other methods, are brought to the energy of interest and interact with the vacuum near-field of an empty optical resonator. As a result, an entangled electron-photon state is formed. Depending on the interaction parameters, a wide range of states can be generated (Fig.~\ref{fig:abstract}, tables). The effect responsible for the electron-light coupling is the near-field Cherenkov radiation \cite{cerenkov_visible_1934, tamm1937coherent, ginzburg1940quantum, tamm_general_1960, ginzburg1996radiation, adiv_observation_2023}. In the presence of a diffraction grating (Fig.~\ref{fig:abstract}), this effect can also be considered the Smith-Purcell radiation into the cavity \cite{doucas_first_1992, remez_spectral_2017}. Here, we focus our attention on the case of a cavity mode initially empty or populated with photon Fock states, although the approach can be applied to more general cases.

Flying past the cavity, an electron can emit photons into it or absorb photons from it. For the interaction to be efficient, it is necessary to fulfill the phase-matching condition (Fig.~\ref{fig:dispersion}a), so the change of electron momentum should be matched with the photon momentum (so-called \textit{direct phase-matching}). The dispersion curves of slow and fast electrons are shown schematically in Fig.~\ref{fig:dispersion}b. In case of fast electrons, the electron dispersion is almost linear, which results in the generation of photons with Poissonian statistics \cite{kfir_entanglements_2019}. In contrast, the dispersion curve of slow electrons is parabolic, which prevents the consecutive multiple emission of photons with the same frequency due to the phase mismatch (Fig.~\ref{fig:dispersion}b, multi-mode inset). This is attributed to the quantum recoil effect \cite{huang_quantum_2023}, where the linear dispersion implies weak recoil and the parabolic a non-negligible one: To compensate the phase-mismatch appearing after the emission of the first photon, the next photon needs to have a slightly different frequency. *

To visualize the phase-matching condition, it is convenient to use the coordinates ($\Delta \Omega$, $\omega$) versus ($\Delta k$, $\varkappa$). Here, $\Delta \Omega$ and $\Delta k$ are the changes of the electron frequency and momentum after the photon emission, whereas $\omega$ and $\varkappa$ are the photon angular frequency and momentum, respectively. Thus, the phase-matching condition is represented by the intersection between the photon dispersion curve (orange) and the electron frequency and momentum change curve ($\Delta \Omega$ on $\Delta k$, blue) in Fig.~\ref{fig:dispersion}c. For photon energies much smaller than the kinetic electron energy we can write the energy-momentum conservation as $\Delta \Omega = \omega$ and $v^{photon}_{ph} \equiv \omega / \varkappa = \Delta \Omega / \Delta k \approx v^{electron}_{gr}$, where the equality of the electron group velocity $v^{electron}_{gr}$ and the photon phase velocity $v^{photon}_{ph}$ is often used in the literature as an equivalent phase-matching condition \cite{park_photon-induced_2010}.

For direct phase-matching and normal photon dispersion, a blue shift to higher photon frequencies can be observed because the speed of the electron is reduced, hence only photons with higher frequency (and lower phase velocity) can fulfill the phase-matching condition $v^{electron}_{gr} = v^{photon}_{ph}$ with the slowed down electron (Fig.~\ref{fig:dispersion}c).

An interesting case is the interaction of free electrons with an effectively single-mode cavity (Fig.~\ref{fig:dispersion}b, single-mode inset). This means that the electron can be phase-matched to one resonant frequency and not to another following the photon emission. In this case, the recoil effect results in the recoil-induced shaping of the quantum state of an electron and a photon. In the case of an effectively single-mode cavity with a phase-matched single-photon transition, the recoil forbids the generation of higher photon states, resulting in antibunching (Fig.~\ref{fig:dispersion}(b,f)) and creation of a single-photon state. We call this effect recoil-induced antibunching, a particular case of the recoil-induced shaping. Later in the article we quantitatively describe the single-mode cavity case.

It is also possible to use the diffraction grating vector $\vec{G}$ to fulfill the phase-matching condition \cite{doucas_first_1992, remez_spectral_2017, karnieli_jaynes-cummings_2023}, which is called quasi-phase-matching (QPM). QPM can change the shift sign of the emitted photon frequency depending on the group velocity relation. Recoil with QPM, when the group velocity of the electron is larger than the \textit{group} velocity of the photon $v^{electron}_{gr}>v^{photon}_{gr}$, results in a blue shift of the generated photon (Fig.~\ref{fig:dispersion}d), while recoil with QPM and $v^{electron}_{gr}<v^{photon}_{gr}$ results in a red shift of the generated photon (Fig.~\ref{fig:dispersion}e). Note that to fulfil the phase-matching condition the photon \textit{phase} velocity needs to satisfy the QPM condition $\Delta k - \varkappa - |\vec{G}| = 0$ or $\omega(1/v^{electron}_{gr}-1/v^{photon}_{ph})-|\vec{G}| = 0$. We can also phase-match two-photon transitions or even higher order ones (Fig.~\ref{fig:dispersion}f) using QPM, which can result in the creation of shaped electron-photon squeezed vacuum state. 

\section{Hamiltonian derivation}\label{sec:Supp_hamiltonian}
\subsection{Dirac field of the electron}

We consider an electron propagating with the momentum $\textbf{p} = \{0, 0, p\}$ along the z-axis. 
The quantized Dirac field of the electron reads \cite{cohen-tannoudji_photons_1997}:
\begin{equation}
  \hat{\Psi}(z) = \tilde{v}^{-1/2} \sum_{p,\sigma} u_{p,\sigma}(z) \hat{c}_{p,\sigma},
\end{equation}
where $\hat{c}_{p,\sigma}$ is the annihilation operator of the electron with the momentum $p$ and spin $\sigma=+\ \text{or}\ -$, and $\tilde{v}$ is a normalization constant. The function $u_{p,\sigma}(r)$ is given by the following expression in the plane-wave basis:
\begin{equation}
  u_{p,\sigma}(z) = 
  \begin{bmatrix}
    C_p I & -S_p \sigma_z\\
    S_p \sigma_z & C_p I
  \end{bmatrix}
  \begin{bmatrix}
    A_{\sigma} \\
    
    0
  \end{bmatrix}
  e^{i p z/\hbar},
\end{equation}
where $C_p = \cos(\theta_p/2)$, $S_p = \sin(\theta_p/2)$, $\theta_p = \arctan(\dfrac{p}{mc})$,
\begin{equation}
  A_{+} = 
  \begin{bmatrix}
  1\\0  
  \end{bmatrix}, \ \ \ \ \ 
    A_{-} = 
  \begin{bmatrix}
  0\\1  
  \end{bmatrix}.
\end{equation}
From this we obtain
\begin{equation}
  u_{p,+}(z) = 
  \begin{bmatrix}
    C_p\\
    0\\
    S_p\\
    0 
  \end{bmatrix}
  e^{i p z/\hbar}, \ \ \ \ \ 
  u_{p,-}(z) = 
  \begin{bmatrix}
    0\\
    C_p\\
    0\\
    -S_p 
  \end{bmatrix}
  e^{i p z/\hbar}.
\end{equation}

Let us calculate the current: 
\begin{equation}
  \hat{j}_n(r) = qc \widetilde{\hat{\Psi}}(r)^\dagger \alpha_n \hat{\Psi}(r) = qc/\tilde{v} \sum_{p_1, p_2, \sigma_1, \sigma_2} \hat{c}_{p_1, \sigma_1}^{\dagger}\hat{c}_{p_2, \sigma_2} \widetilde{u}^*_{p_1, \sigma_1} \alpha_n u_{p_2, \sigma_2}.
\end{equation}

$\widetilde{\Psi}$ is the transpose of $\Psi$ and $\alpha_n$ is the $n^{th}$ Dirac matrix.

First, we calculate the $\alpha_n u_{p_2, \sigma_2}$ terms ($X_+ \equiv \alpha_x u_{p_2, +}$ and so on):
\begin{equation}
  X_{+} =
  \begin{bmatrix}
    0 & 0 & 0 & 1\\
    0 & 0 & 1 & 0\\
    0 & 1 & 0 & 0\\
    1 & 0 & 0 & 0\\
  \end{bmatrix}
  \begin{bmatrix}
    C_{p_2}\\
    0\\
    S_{p_2}\\
    0
  \end{bmatrix}
  e^{ip_2z/\hbar}
  =
  \begin{bmatrix}
    0\\
    S_{p_2}\\
    0\\
    C_{p_2}
  \end{bmatrix}
  e^{ip_2z/\hbar},
\end{equation}
\begin{equation}
  X_{-} =
  \begin{bmatrix}
    0 & 0 & 0 & 1\\
    0 & 0 & 1 & 0\\
    0 & 1 & 0 & 0\\
    1 & 0 & 0 & 0\\
  \end{bmatrix}
  \begin{bmatrix}
    0\\
    C_{p_2}\\
    0\\
    -S_{p_2}
  \end{bmatrix}
  e^{ip_2z/\hbar}
  =
  \begin{bmatrix}
    -S_{p_2}\\
    0\\
    C_{p_2}\\
    0
  \end{bmatrix}
  e^{ip_2z/\hbar},
\end{equation}

\begin{equation}
  Y_{+} =
  \begin{bmatrix}
    0 & 0 & 0 & -i\\
    0 & 0 & i & 0\\
    0 & -i & 0 & 0\\
    i & 0 & 0 & 0\\
  \end{bmatrix}
  \begin{bmatrix}
    C_{p_2}\\
    0\\
    S_{p_2}\\
    0
  \end{bmatrix}
  e^{ip_2z/\hbar}
  =
  \begin{bmatrix}
    0\\
    i S_{p_2}\\
    0\\
    i C_{p_2}
  \end{bmatrix}
  e^{ip_2z/\hbar},
\end{equation}
\begin{equation}
  Y_{-} =
  \begin{bmatrix}
    0 & 0 & 0 & -i\\
    0 & 0 & i & 0\\
    0 & -i & 0 & 0\\
    i & 0 & 0 & 0\\
  \end{bmatrix}
  \begin{bmatrix}
    0\\
    C_{p_2}\\
    0\\
    -S_{p_2}
  \end{bmatrix}
  e^{ip_2z/\hbar}
  =
  \begin{bmatrix}
    iS_{p_2}\\
    0\\
    -iC_{p_2}\\
    0
  \end{bmatrix}
  e^{ip_2z/\hbar},
\end{equation}

\begin{equation}
  Z_{+} =
  \begin{bmatrix}
    0 & 0 & 1 & 0\\
    0 & 0 & 0 & -1\\
    1 & 0 & 0 & 0\\
    0 & -1 & 0 & 0\\
  \end{bmatrix}
  \begin{bmatrix}
    C_{p_2}\\
    0\\
    S_{p_2}\\
    0
  \end{bmatrix}
  e^{ip_2z/\hbar}
  =
  \begin{bmatrix}
    S_{p_2}\\
    0\\
    C_{p_2}\\
    0
  \end{bmatrix}
  e^{ip_2z/\hbar},
\end{equation}
\begin{equation}
  Z_{-} =
  \begin{bmatrix}
    0 & 0 & 1 & 0\\
    0 & 0 & 0 & -1\\
    1 & 0 & 0 & 0\\
    0 & -1 & 0 & 0\\
  \end{bmatrix}
  \begin{bmatrix}
    0\\
    C_{p_2}\\
    0\\
    -S_{p_2}
  \end{bmatrix}
  e^{ip_2z/\hbar}
  =
  \begin{bmatrix}
    0\\
    S_{p_2}\\
    0\\
    -C_{p_2}
  \end{bmatrix}
  e^{ip_2z/\hbar}.
\end{equation}

Using these expressions, one can obtain the components of the current density:
\begin{equation*}
\begin{gathered}
  \hat{j}_x = qc/\tilde{v} \sum_{p_1, p_2} (C_{p_1}S_{p_2} - S_{p_1}C_{p_2})(\hat{c}^{\dagger}_{p_1,-} \hat{c}_{p_2,+} - \hat{c}^{\dagger}_{p_1,+} \hat{c}_{p_2,-}) e^{i(p_2 - p_1)z/\hbar} =\\= qc/\tilde{v} \sum_{p_1, p_2} S_{p_2 - p_1} (\hat{c}^{\dagger}_{p_1,-} \hat{c}_{p_2,+} - \hat{c}^{\dagger}_{p_1,+} \hat{c}_{p_2,-}) e^{i(p_2 - p_1)z/\hbar}, \\
  \hat{j}_y = iqc/\tilde{v} \sum_{p_1, p_2} (C_{p_1}S_{p_2} - S_{p_1}C_{p_2})(\hat{c}^{\dagger}_{p_1,-} \hat{c}_{p_2,+} + \hat{c}^{\dagger}_{p_1,+} \hat{c}_{p_2,-}) e^{i(p_2 - p_1)z/\hbar} =\\= iqc/\tilde{v} \sum_{p_1, p_2} S_{p_2 - p_1} (\hat{c}^{\dagger}_{p_1,-} \hat{c}_{p_2,+} + \hat{c}^{\dagger}_{p_1,+} \hat{c}_{p_2,-}) e^{i(p_2 - p_1)z/\hbar}, \\
\hat{j}_z = qc/\tilde{v} \sum_{p_1, p_2} (C_{p_1}S_{p_2} + S_{p_1}C_{p_2})(\hat{c}^{\dagger}_{p_1,+} \hat{c}_{p_2,+} + \hat{c}^{\dagger}_{p_1,-} \hat{c}_{p_2,-}) e^{i(p_2 - p_1)z/\hbar} =\\= qc/\tilde{v} \sum_{p_1, p_2} S_{p_2 + p_1} (\hat{c}^{\dagger}_{p_1,+} \hat{c}_{p_2,+} + \hat{c}^{\dagger}_{p_1,-} \hat{c}_{p_2,-}) e^{i(p_2 - p_1)z/\hbar}.
\end{gathered}
\end{equation*}

We consider the interaction of the electron with the $z$-component of the cavity mode (quasi-TM$_{00}$, for example). So, the $z$-component of this mode quantized vector potential is:
\begin{equation}
  \hat{A}_z(\mathbf{r}) = \hat{A}^{\perp}_{z}(x,y) \hat{A}^{\|}_z(z) = \hat{A}^{\perp}_{z}(x,y) \left( \sum_k \mathcal{A}(k) \hat{a}_k e^{ikz} + \mathcal{A}^*(k) \hat{a}^{\dagger}_k e^{-ikz} \right).
\end{equation}

The interaction Hamiltonian reads:
\begin{equation}
\begin{gathered}
  V = -\int \mathrm{d}^3 \mathbf{r} \ \hat{\mathbf{j}}(\mathbf{r}) \hat{\mathbf{A}}(\mathbf{r}) = -\eta \int \mathrm{d} z \hat{j}^{\|}_z(z) \hat{A}^{\|}_z(z) = \\ = - \tilde{\eta} \sum_{p_1, p_2, k} \int \mathrm{d}z~ S_{p_1 + p_2} (\hat{c}^{\dagger}_{p_1,+} \hat{c}_{p_2,+} + \hat{c}^{\dagger}_{p_1,-} \hat{c}_{p_2,-}) e^{i(p_2 - p_1)z/\hbar} \big[\mathcal{A}(k) \hat{a}_k e^{ikz} + \mathcal{A}^*(k) \hat{a}^{\dagger}_k e^{-ikz}\big],
\end{gathered}
\end{equation}
where $\eta=\int \mathrm{d}x\mathrm{d}y\ \hat{A}^{\perp}_{z}(x,y) \ \hat{j}^{\perp}_z(x,y)$ is the transverse overlap. We can assume $\hat{j}_z(\mathbf{r})=\hat{j}^{\|}_z(z)\hat{j}^{\perp}_z(x,y)=\hat{j}^{\|}_z(z)\delta(x)\delta(y)$, so $\eta=\hat{A}^{\perp}_{z}(0,0)$ and $\tilde{\eta} = \hat{A}^{\perp}_{z}(0,0) qc\tilde{v}^{-1/3}$. Here the remaining normalization constant along the propagation direction is $\tilde{v}^{-1/3}=1/L$, where $L$ is the cavity length.

Let's assume that the electron spin $\sigma = +$, therefore
\begin{equation}
\begin{gathered}
V = -\tilde{\eta} \sum_{p_1, p_2, k} \int \mathrm{d}z~ S_{p_1 + p_2}\bigg[\mathcal{A}(k) \hat{c}^{\dagger}_{p_1} \hat{c}_{p_2}\hat{a}_k e^{i(p_2 - p_1)z/\hbar + ikz} + \mathcal{A}^*(k) \hat{c}^{\dagger}_{p_1} \hat{c}_{p_2}\hat{a}^{\dagger}_k e^{i(p_2 - p_1)z/\hbar - ikz}\bigg] = \\ = \sum_{p_1,p_2,k}V_{p_1p_2k}\hat{c}^{\dagger}_{p_1} \hat{c}_{p_2}\hat{a}_k + \sum_{p_1,p_2,k}W_{p_1p_2k}\hat{c}^{\dagger}_{p_1} \hat{c}_{p_2}\hat{a}_k^{\dagger}.
\end{gathered}
\end{equation}
The Hamiltonian of the free fields (electrons and photons) is:
\begin{equation}
  \hat{H}_0 = \sum_p E_p \hat{c}_p^{\dagger}\hat{c}_p + \sum_{k'} \hbar \omega_{k'} \hat{a}_{k'}^{\dagger} \hat{a}_{k'},
\end{equation}
where $E_p = \sqrt{p^2 c^2 + m^2 c^4}$.
The Hamiltonian in the interaction picture:
\begin{equation}
  H_I = e^{i H_0 t /\hbar} V e^{-i H_0 t /\hbar} = V + i t/\hbar [H_0, V] + \frac{(it/\hbar)^2}{2!} [H_0, [H_0, V]] + \ ...
  \label{eq:expon_comm}
\end{equation}
Let's calculate the commutators:
\begin{equation*}
  [\hat{c}_p^{\dagger} \hat{c}_p, \hat{c}^{\dagger}_{p_1}] = \hat{c}_p^{\dagger} \hat{c}_p \hat{c}^{\dagger}_{p_1} - \hat{c}^{\dagger}_{p_1}\hat{c}_p^{\dagger} \hat{c}_p = \hat{c}_p^{\dagger} \hat{c}_p \hat{c}^{\dagger}_{p_1} + \hat{c}_p^{\dagger} \hat{c}^{\dagger}_{p_1} \hat{c}_p = \hat{c}_p^{\dagger} \{ \hat{c}_p, \hat{c}^{\dagger}_{p_1} \} = \delta_{p, p_1} \hat{c}_p^{\dagger},
\end{equation*}
\begin{equation*}
  [\hat{c}_p^{\dagger} \hat{c}_p, \hat{c}_{p_1}] = \hat{c}_p^{\dagger} \hat{c}_p \hat{c}_{p_1} - \hat{c}_{p_1}\hat{c}_p^{\dagger} \hat{c}_p = -\hat{c}_p^{\dagger} \hat{c}_{p_1} \hat{c}_p- \hat{c}_{p_1}\hat{c}_p^{\dagger} \hat{c}_p = - \{ \hat{c}_{p_1}, \hat{c}^{\dagger}_{p} \} \hat{c}_p = -\delta_{p, p_1} \hat{c}_p,
\end{equation*}
\begin{equation*}
  [\hat{c}_p^{\dagger} \hat{c}_p, \hat{c}^{\dagger}_{p_1} \hat{c}_{p_2}\hat{a}_k] = [\hat{c}_p^{\dagger} \hat{c}_p, \hat{c}^{\dagger}_{p_1}] \hat{c}_{p_2}\hat{a}_k + \hat{c}^{\dagger}_{p_1}[\hat{c}_p^{\dagger} \hat{c}_p, \hat{c}_{p_2}]\hat{a}_k = \delta_{p, p_1}\hat{c}^{\dagger}_p \hat{c}_{p_2}\hat{a}_k - \delta_{p,p_2} \hat{c}^{\dagger}_{p_1} \hat{c}_p \hat{a}_k = (\delta_{p,p_1} - \delta_{p,p_2}) \hat{c}^{\dagger}_{p_1} \hat{c}_{p_2} \hat{a}_k,
\end{equation*}
\begin{equation*}
  [\hat{a}_{k'}^{\dagger} \hat{a}_{k'}, \hat{c}^{\dagger}_{p_1} \hat{c}_{p_2}\hat{a}_{k}] = \hat{c}^{\dagger}_{p_1} \hat{c}_{p_2}[\hat{a}_{k'}^{\dagger} \hat{a}_{k'}, \hat{a}_{k}] = -\delta_{k,k'} \hat{c}^{\dagger}_{p_1} \hat{c}_{p_2} \hat{a}_{k'},
\end{equation*}
\begin{equation*}
  [\hat{a}_{k'}^{\dagger} \hat{a}_{k'}, \hat{c}^{\dagger}_{p_1} \hat{c}_{p_2}\hat{a}_{k}^{\dagger}] = \hat{c}^{\dagger}_{p_1} \hat{c}_{p_2}[\hat{a}_{k'}^{\dagger} \hat{a}_{k'}, \hat{a}_{k}^{\dagger}] = \delta_{k,k'} \hat{c}^{\dagger}_{p_1} \hat{c}_{p_2} \hat{a}_{k'}^{\dagger}.
\end{equation*}
As a result, the first commutator in (\ref{eq:expon_comm}) is:
\begin{multline}
  it [H_0, V] = it \bigg[\sum_{p}E_p \hat{c}_p^{\dagger} \hat{c}_p, \sum_{p_1,p_2,k}V_{p_1p_2k}\hat{c}^{\dagger}_{p_1} \hat{c}_{p_2}\hat{a}_k + \sum_{p_1,p_2,k}W_{p_1p_2k}\hat{c}^{\dagger}_{p_1} \hat{c}_{p_2}\hat{a}_k^{\dagger} \bigg] +\\+ it \bigg[\sum_{k'}\hbar \omega_{k'} \hat{a}_{k'}^{\dagger} \hat{a}_{k'}, \sum_{p_1,p_2,k}V_{p_1p_2k}\hat{c}^{\dagger}_{p_1} \hat{c}_{p_2}\hat{a}_k + \sum_{p_1,p_2,k}W_{p_1p_2k}\hat{c}^{\dagger}_{p_1} \hat{c}_{p_2}\hat{a}_k^{\dagger} \bigg] =\\=
  \sum_{p_1,p_2,k} V_{p_1p_2k} it(E_{p_1} - E_{p_2}) \hat{c}^{\dagger}_{p_1} \hat{c}_{p_2}\hat{a}_k + \sum_{p_1,p_2,k} W_{p_1p_2k} it(E_{p_1} - E_{p_2}) \hat{c}^{\dagger}_{p_1} \hat{c}_{p_2}\hat{a}_k^{\dagger} +\\+ \sum_{p_1,p_2,k} V_{p_1p_2k} (-it) \hbar\omega_{k} \hat{c}^{\dagger}_{p_1} \hat{c}_{p_2}\hat{a}_k + \sum_{p_1,p_2,k} W_{p_1p_2k} (it) \hbar\omega_{k} \hat{c}^{\dagger}_{p_1} \hat{c}_{p_2}\hat{a}_k =\\=
  it\sum_{p_1,p_2,k} V_{p_1p_2k} (E_{p_1} - E_{p_2} - \hbar \omega_k) \hat{c}^{\dagger}_{p_1} \hat{c}_{p_2}\hat{a}_k + it\sum_{p_1,p_2,k} W_{p_1p_2k} (E_{p_1} - E_{p_2} + \hbar \omega_k) \hat{c}^{\dagger}_{p_1} \hat{c}_{p_2}\hat{a}_k^{\dagger}.
\end{multline}
Thus, each commutator in (\ref{eq:expon_comm}) gives an exponential series term. After the summation of the series we get:
\begin{equation}
  \begin{aligned}
   H_I = \int \mathrm{d}z \sum_{p_1,p_2,k} g_{p_1, p_2, k} e^{i (\frac{p_2 - p_1}{\hbar} + k) z} e^{-i (\frac{E_{p_2} - E_{p_1}}{\hbar} + \omega_k) t} \hat{c}^{\dagger}_{p_1} \hat{c}_{p_2}\hat{a}_k \\+ \int \mathrm{d}z \sum_{p_1,p_2,k} g^*_{p_1, p_2, k} e^{i (\frac{p_2 - p_1}{\hbar} - k) z} e^{-i (\frac{E_{p_2} - E_{p_1}}{\hbar} - \omega_k) t} \hat{c}^{\dagger}_{p_1} \hat{c}_{p_2}\hat{a}_k^{\dagger},  
  \end{aligned}
\end{equation}
where $g_{p_1, p_2, k} = -\tilde{\eta} S_{p_1 + p_2} \mathcal{A}_k$. For non-relativistic electrons $S_{p_1 + p_2} \approx \frac{p_1+p_2}{2mc} \approx (v_{p_1}+v_{p_2})/(2c)$, and $g_{p_1, p_2, k}$ can further be simplified to $g_{p_1, k} \approx - \tilde{\eta} v_{p_1} \mathcal{A}_k /c$, if we assume that the electron velocity changes insignificantly ($v_{p_1}+v_{p_2} \approx 2 v_{p_1}$). 

Let's change the notation for further convenience: $\Omega(k)$ and $k$ represent the electron energy divided by $\hbar$ and its momentum, while $\omega (\varkappa)$ and $\varkappa$ stand for the photon energy divided by $\hbar$ and its momentum. Also, we rearrange the constants, so that the interaction Hamiltonian is:
\begin{equation}
\hat{H}_{int} = i \hbar \sum_{k_1,k_2,\varkappa} \int \mathrm{d} z \tilde{g}_{k_1, k_2, \varkappa} e^{i(k_1-k_2-\varkappa)z} e^{-i(\Omega_{k_1}-\Omega_{k_2}-\omega_{\varkappa})t} \hat{c}_{k_2}^{\dagger}(t) \hat{c}_{k_1}(t) \hat{a}^{\dagger}_{\varkappa}(t) + h.c.
\label{eq:hamilt_initial_discrete_k}
\end{equation}
And the corresponding Schrödinger equation for the electron-photon wavefunction $|\psi(t)\rangle$ is
\begin{equation}
i \hbar \frac{\mathrm{d}}{\mathrm{d} t}|\psi(t)\rangle=\hat{H}_{int}(t)|\psi(t)\rangle.
\end{equation}

\subsection{Hamiltonian Fourier transform}
Since the interaction of free electrons with the cavity is a long macroscopic process, we can make use of the approaches developed in quantum nonlinear optics for the description of parametric down-conversion (PDC) \cite{klyshko_photons_2017, sharapova_properties_2020}. We can write the continualized version of the Hamiltonian (for brevity $\Omega_{k_1} = \Omega_1, \ \Omega_{k_2} = \Omega_2$ and $\omega_{\varkappa} = \omega$): 
\begin{equation}
  \hat{H}_{int} = i \hbar \int \mathrm{d} z \mathrm{d} k_1 \mathrm{d} k_2 \mathrm{d} \varkappa \Gamma e^{i(k_1-k_2-\varkappa)z} e^{-i(\Omega_{1}-\Omega_{2}-\omega)t} \hat{c}^{\dagger}(k_2, t) \hat{c}(k_1, t) \hat{a}^{\dagger}(\varkappa, t) + h.c.
  \label{eq:hamiltonian_continuum}
\end{equation}
The corresponding Heisenberg equations $\frac{\mathrm{d}}{\mathrm{d} t} \hat{A} = { \frac{i}{\hbar}} [\hat{H},\hat{A}]$ for the photon and electron operators are:
\begin{equation}
\begin{aligned}
\frac{\mathrm{d} \hat{a}(\varkappa, t)}{\mathrm{d} t} = &\int \mathrm{d} z \mathrm{d} k_1 \mathrm{d} k_2 \Gamma e^{i(k_1-k_2-\varkappa)z} e^{-i(\Omega_{1}-\Omega_{2}-\omega)t} \hat{c}^{\dagger}(k_2, t) \hat{c}(k_1, t), \\
\frac{\mathrm{d} \hat{c}(p, t)}{\mathrm{d} t} = &\int \mathrm{d} z \mathrm{d} k_1 \mathrm{d} \varkappa \Gamma e^{i(k_1-p-\varkappa)z} e^{-i(\Omega_{1}-\Omega_{p}-\omega)t} \hat{a}^{\dagger}(\varkappa, t) \hat{c}(k_1, t) - \\
- &\int \mathrm{d} z \mathrm{d} k_2 \mathrm{d} \varkappa \Gamma^* e^{-i(p-k_2-\varkappa)z} e^{i(\Omega_{p}-\Omega_{2}-\omega)t} \hat{a}(\varkappa, t) \hat{c}(k_2, t).
  \label{eq:heisenberg}
\end{aligned}
\end{equation}
The next step is to take the Fourier transform (FT) of operators to go from the $(k, t)$ to the $(\Omega, L)$ representation:
\begin{gather}
\hat{c}^{\dagger}(k, t) \xrightarrow{FT} \hat{c}^{\dagger}(\Omega, L).
\label{eq:fourier_c}
\end{gather}
First, we introduce the fast varying components of the operators:
\begin{equation}
\begin{gathered}
\bar{\hat{c}}^{\dagger}(k_2, t) = e^{i\Omega_{2}t} \hat{c}^{\dagger}(k_2, t), \\
\bar{\hat{c}}(k_1, t) = e^{-i\Omega_{1}t} \hat{c}(k_1, t), \\
\bar{\hat{a}}^{\dagger}(\varkappa, t) = e^{i\omega t} \hat{a}^{\dagger}(\varkappa, t), \\
\bar{\hat{a}}(\varkappa, t) = e^{-i\omega t} \hat{a}(\varkappa, t).
\label{eq:fast_slow}
\end{gathered}
\end{equation}
The Fourier transform of these operators:
\begin{equation}
\begin{gathered}
\bar{\hat{c}}(k, t) = \frac{1}{2\pi} \int \bar{\hat{c}}(\tilde{\Omega}, \xi) e^{-i\tilde{\Omega}t} e^{-i k\xi} \mathrm{d} \tilde{\Omega} \mathrm{d} \xi = \frac{1}{2\pi} \int {\hat{c}}(\tilde{\Omega}, \xi) e^{i k(\tilde{\Omega})\xi} e^{-i\tilde{\Omega}t} e^{-i k\xi} \mathrm{d} \tilde{\Omega} \mathrm{d} \xi, \\
\bar{\hat{c}}^{\dagger}(k, t) = \frac{1}{2\pi} \int \hat{c}^{\dagger}(\tilde{\Omega}, \xi) e^{-i k(\tilde{\Omega})\xi} e^{i\tilde{\Omega}t} e^{i k\xi} \mathrm{d} \tilde{\Omega} \mathrm{d} \xi,\\
\bar{\hat{a}}^{\dagger}(\varkappa, t) = \frac{1}{2\pi} \int \hat{a}^{\dagger}(\tilde{\omega}, \xi) e^{-i \varkappa(\tilde{\omega})\xi} e^{i\tilde{\omega}t} e^{i \varkappa\xi} \mathrm{d} \tilde{\omega} \mathrm{d} \xi,\\
\bar{\hat{a}}(\varkappa, t) = \frac{1}{2\pi} \int \hat{a}(\tilde{\omega}, \xi) e^{i \varkappa(\tilde{\omega})\xi} e^{-i\tilde{\omega}t} e^{-i \varkappa\xi} \mathrm{d} \tilde{\omega} \mathrm{d} \xi.
\label{eq:fourier_operators}
\end{gathered}
\end{equation}

Integrating the Heisenberg equation (\ref{eq:heisenberg}) over time $t$ from $\tau_0=-\infty$ to $\tau=\infty$ and substituting (\ref{eq:fourier_operators}), we get for the photon operator:
\begin{equation}
\begin{gathered}
\hat{a}(\varkappa, \tau)-\hat{a}(\varkappa, \tau_0) = \int \mathrm{d}t \mathrm{d} z \mathrm{d} k_1 \mathrm{d} k_2 \Gamma e^{i(k_1-k_2-\varkappa)z} e^{i\omega t} \times \\
\times \frac{1}{2\pi} \int \hat{c}^{\dagger}(\tilde{\Omega}_2, \xi_2) e^{-i k(\tilde{\Omega}_2)\xi_2} e^{i\tilde{\Omega}_2t} e^{i k_2\xi_2} \mathrm{d} \tilde{\Omega}_2 \mathrm{d} \xi_2 \times \frac{1}{2\pi} \int {\hat{c}}(\tilde{\Omega}_1, \xi_1) e^{i k(\tilde{\Omega}_1)\xi_1} e^{-i\tilde{\Omega}_1t} e^{-i k_1\xi_1} \mathrm{d} \tilde{\Omega}_1 \mathrm{d} \xi_1.
\label{eq:photon_integrals}
\end{gathered}
\end{equation}
We note that $k(\tilde{\Omega}_2)$ is a function of $\tilde{\Omega}_2$, while $k_2$ is the integration variable. For the electron operator we get:
\begin{equation}
\begin{gathered}
\hat{c}(p, \tau)-\hat{c}(p, \tau_0) = \int \mathrm{d}t \mathrm{d} z \mathrm{d} k_1 \mathrm{d} \varkappa \Gamma e^{i(k_1-p-\varkappa)z} e^{i\Omega_p t} \times \\
\times \frac{1}{2\pi} \int \hat{a}^{\dagger}(\tilde{\omega}, \xi) e^{-i \varkappa(\tilde{\omega})\xi} e^{i\tilde{\omega}t} e^{i \varkappa\xi} \mathrm{d} \tilde{\omega} \mathrm{d} \xi \times \frac{1}{2\pi} \int {\hat{c}}(\tilde{\Omega}_1, \xi_1) e^{i k(\tilde{\Omega}_1)\xi_1} e^{-i\tilde{\Omega}_1t} e^{-i k_1\xi_1} \mathrm{d} \tilde{\Omega}_1 \mathrm{d} \xi_1 - \\
- \int \mathrm{d}t \mathrm{d} z \mathrm{d} k_2 \mathrm{d} \varkappa \Gamma^* e^{-i(p-k_2-\varkappa)z} e^{i\Omega_p t} \times \\
\times \frac{1}{2\pi} \int \hat{a}(\tilde{\omega}, \xi) e^{i \varkappa(\tilde{\omega})\xi} e^{-i\tilde{\omega}t} e^{-i \varkappa\xi} \mathrm{d} \tilde{\omega} \mathrm{d} \xi 
\times \frac{1}{2\pi} \int \hat{c}(\tilde{\Omega}_2, \xi_2) e^{i k(\tilde{\Omega}_2)\xi_2} e^{-i\tilde{\Omega}_2t} e^{-i k_2\xi_2} \mathrm{d} \tilde{\Omega}_2 \mathrm{d} \xi_2.
\label{eq:electron_integrals}
\end{gathered}
\end{equation}
Similar to above, here $\varkappa(\tilde{\omega})$ is a function of $\tilde{\omega}$, while $\varkappa$ is the integration variable.

Integration in (\ref{eq:photon_integrals}) and (\ref{eq:electron_integrals}) leads to the $\delta$-functions of the form
\begin{equation}
   \frac{1}{2\pi} \int_{-\infty}^{\infty} \mathrm{d} t~ e^{-i(\tilde{\Omega}_{1}-\tilde{\Omega}_{2}-{\omega})t} = \delta(\tilde{\Omega}_{1}-\tilde{\Omega}_{2}-{\omega}),
\end{equation}
which determine the electron energy/frequency ladder of levels for this problem and the perfect entanglement between electron energy value and photon number state. Also we get other $\delta$-functions of the form
\begin{equation}
   \frac{1}{2\pi} \int_{-\infty}^{\infty} \mathrm{d} k_1~ e^{ik_1(z-\xi_1)} = \delta(z-\xi_1),
\end{equation}
so that from 8 integrals in each equation (\ref{eq:photon_integrals}, \ref{eq:electron_integrals}) only 2 survive.

As a result, from (\ref{eq:photon_integrals}) and (\ref{eq:electron_integrals}) we get 
\begin{equation}
\begin{aligned}
\hat{a}(\varkappa, \tau)-\hat{a}(\varkappa, \tau_0) &= 2\pi \int_{0}^{L} \mathrm{d} z \int \mathrm{d} \tilde{\Omega}_1 \Gamma e^{i(k(\tilde{\Omega}_1)-k(\tilde{\Omega}_1-\omega)-\varkappa)z} \hat{c}^{\dagger}(\tilde{\Omega}_1-\omega, z) {\hat{c}}(\tilde{\Omega}_1, z),\\
\hat{c}(p, \tau)-\hat{c}(p, \tau_0) &= 2\pi \int_{0}^{L} \mathrm{d} z \int \mathrm{d} \tilde{\omega} \Gamma e^{i(k(\Omega_p+\tilde{\omega})-p-\varkappa(\tilde{\omega}))z} \hat{a}^{\dagger}(\tilde{\omega}, z) {\hat{c}}(\Omega_p+\tilde{\omega}, z)-\\
&- 2\pi \int \mathrm{d} z \mathrm{d} \tilde{\omega} \Gamma^* e^{-i(p-k(\Omega_p-\tilde{\omega})-\varkappa(\tilde{\omega}))z} \hat{a}(\tilde{\omega}, z) {\hat{c}}(\Omega_p-\tilde{\omega}, z),
\label{eq:integrals_after}
\end{aligned}
\end{equation}
where $L$ is the length of interaction.

The connection between the boundary conditions in $(k, t)$ and $(\Omega, L)$ representations can be written as \cite{klyshko_photons_2017, sharapova_properties_2020}: 
\begin{equation}
\begin{aligned}
  \hat{a}(\varkappa, \tau)=u\hat{a}(\omega,L),& \ \ \hat{a}(\varkappa, \tau_0) = u\hat{a}(\omega,0), \\ \hat{c}(p, \tau) = v \hat{c}(\Omega, L),& \ \ \hat{c}(p, \tau_0) = v\hat{c}(\Omega, 0),
  \label{eq:boundary}
\end{aligned}
\end{equation}
where $u$ and $v$ are the photon and electron group velocities, respectively. From the boundary conditions we also get $[\hat{a}({\omega}_1, L), \hat{a}^{\dagger}({\omega}_2, L)]=\frac{1}{u^2} [\hat{a}(\varkappa_1, \tau), \hat{a}^{\dagger}(\varkappa_2, \tau)]=\frac{1}{u^2} \delta(\varkappa_1-\varkappa_2) = \frac{1}{u} \delta(\omega_1-\omega_2)$.

Differentiation of the both sides of Eq.~\ref{eq:integrals_after} with respect to $L$ and variable reassignment lead, finally, to the Heisenberg equations in the $(\Omega, L)$ representation:
\begin{equation}
\begin{aligned}
\frac{\mathrm{d} \hat{a}(\omega, L)}{\mathrm{d} L} = &\frac{2\pi}{u} \int \mathrm{d} \tilde{\Omega} \Gamma e^{i(k(\tilde{\Omega})-k(\tilde{\Omega}-\omega)-\varkappa(\omega))L} \hat{c}^{\dagger}(\tilde{\Omega}-\omega, L) {\hat{c}}(\tilde{\Omega}, L),\\
\frac{\mathrm{d} \hat{c}(\Omega, L)}{\mathrm{d} L} = &\frac{2\pi}{v} \int \mathrm{d} \tilde{\omega} \Gamma e^{i(k({\Omega}+\tilde{\omega})-k(\Omega)-\varkappa(\tilde{\omega}))L} \hat{a}^{\dagger}(\tilde{\omega}, L) {\hat{c}}({\Omega}+\tilde{\omega}, L) - \\
- &\frac{2\pi}{v} \int \mathrm{d} \tilde{\omega} \Gamma^* e^{-i(k(\Omega)-k({\Omega}-\tilde{\omega})-\varkappa(\tilde{\omega}))L} \hat{a}(\tilde{\omega}, L) {\hat{c}}({\Omega}-\tilde{\omega}, L),
  \label{eq:heisenberg_L}
\end{aligned}
\end{equation}
where we changed $\Omega(p) \rightarrow \Omega, \ p \rightarrow k(\Omega)$. Heisenberg equations in this form are useful for the description of the free-electron interaction with many photons (via $\langle \hat{a}^{\dagger} \hat{a} \rangle$, $\langle \hat{c}^{\dagger} \hat{c} \rangle$ and higher moments) — for instance, for the squeezing estimation, non-classical PINEM and harmonics generation with PINEM. We will use these equations here to get the Hamiltonian and Schrödinger equation in the $(\Omega, L)$ representation. 

Using Heisenberg equations $v\frac{\mathrm{d}}{\mathrm{d} L} \hat{c} = { \frac{i}{\hbar}} [\hat{H},\hat{c}]$, we restore the Hamiltonian
\begin{equation}
\begin{aligned}
\hat{H}_{int} (L) &= 2\pi i \hbar v \int \mathrm{d} {\omega} \mathrm{d} \Omega \Gamma e^{i(k(\Omega)-k(\Omega-\omega)-\varkappa({\omega}))L} \hat{c}^{\dagger}({\Omega}-{\omega}, L) \hat{c}({\Omega}, L) \hat{a}^{\dagger}({\omega}, L) - \\
&- 2\pi i \hbar v \int \mathrm{d} {\omega} \mathrm{d} \Omega \Gamma^* e^{-i(k(\Omega+\omega)-k({\Omega})-\varkappa({\omega}))L} \hat{c}^{\dagger}({\Omega}+\omega, L) \hat{c}({\Omega}, L) \hat{a}({\omega}, L).
  \label{eq:Hamiltonian_L}
\end{aligned}
\end{equation}
Here and further we assume that the interaction propagates with the group velocity of the electron $v$. Returning to the discrete form like in the initial Hamiltonian (\ref{eq:hamilt_initial_discrete_k}) and rearranging constants:
\begin{equation}
\begin{aligned}
\hat{H}_{int} (L) &= i \hbar v \sum_{\omega, \Omega} \gamma_{\omega} e^{i(k(\Omega)-k(\Omega-\omega)-\varkappa({\omega}))L} \hat{c}^{\dagger}_{{\Omega}-{\omega}}(L) \hat{c}_{\Omega}(L) \hat{a}^{\dagger}_{\omega}(L) - \\
&- i \hbar v \sum_{\omega, \Omega} \gamma_{\omega}^* e^{-i(k(\Omega+\omega)-k({\Omega})-\varkappa({\omega}))L} \hat{c}^{\dagger}_{{\Omega}+\omega}(L) \hat{c}_{\Omega}(L) \hat{a}_{\omega}(L).
  \label{eq:Hamiltonian_L_discrete}
\end{aligned}
\end{equation}

Now we can write the length-dependent Schrödinger equation $i\hbar v_{} \frac{\mathrm{d}}{\mathrm{d} L}|\psi(L)\rangle = \hat{H}_{int} (L) |\psi(L)\rangle$ to get the spatial evolution of the wavefunction
\begin{equation}
\begin{gathered}
\ket{\psi(L)} = \sum_{\Omega, n_{\omega_1}, n_{\omega_2}, ..., n_{\omega_N}} C_{\Omega, n_{\omega_1}, n_{\omega_2}, ..., n_{\omega_N}}(L) \ket{\Omega, n_{\omega_1}, n_{\omega_2}, ..., n_{\omega_N}},
\label{eq:psi_L}
\end{gathered}
\end{equation}
where $\ket{\Omega, n_{\omega_1}, n_{\omega_2}, ..., n_{\omega_N}}$ is the state of electron with energy $\Omega$ and the cavity with $n_{\omega_j}$ photons in the mode $\omega_j$. 

Considering the interaction of the electron with the multimode state of the cavity, we get the differential equation for the electron-photon wavefunction coefficients (Schrödinger equation in the matrix form):
\begin{equation}
\begin{aligned}
\frac{\mathrm{d}}{\mathrm{d} L} C_{\Omega, n_{\omega_1}, n_{\omega_2}, ..., n_{\omega_N}}(L) = \sum_{j} \ &\tilde{g}_{\rm Qu}^{\omega_j} e^{-i(k(\Omega)-k(\Omega-\omega_j)-\varkappa(\omega_j))L} \sqrt{n_{\omega_j}+1} C_{\Omega-\omega_j, n_{\omega_1}, n_{\omega_2}, ..., n_{\omega_j}+1,..., n_{\omega_N}}(L) \\ &- (\tilde{g}_{\rm Qu}^{\omega_j})^* e^{i(k(\Omega+\omega_j)-k(\Omega)-\varkappa(\omega_j))L} \sqrt{n_{\omega_j}} C_{\Omega+\omega_j, n_{\omega_1}, n_{\omega_2}, ..., n_{\omega_j}-1,..., n_{\omega_N}}(L), 
\end{aligned}
\label{eq:coeff_ode_multimode_supp}
\end{equation}
here $\tilde{g}_{Qu} \equiv -\gamma^* = g_{\rm Qu}/L_{full}$ is a normalized version of the dimensionless coupling parameter $g_{\rm Qu}$ often used in literature \cite{kfir_entanglements_2019}, and $L_{full}$ is the full length of interaction. Comparing the constants with Eq.~\ref{eq:hamilt_initial_discrete_k}, we see that $g_{\rm Qu} = q \hat{A}^{\perp}_{z}(0,0)\mathcal{A}_k \tilde{v}^{-1/3} L^2 /(i\hbar)$. Since $\tilde{v}^{-1/3}=1/L$ and $\mathcal{A}_k \propto 1/\sqrt{L}$ due to normalization, we get $|g_{\rm Qu}^{max}| = q A L /(2\hbar) \propto \sqrt{L}$, which is consistent with the expression in other works \cite{kfir_entanglements_2019}: $|g_{\rm Qu}^{max}| = q E L/(2\hbar \omega) \propto \sqrt{L}$ (here $A$ and $E$ are the vector potential and electric field amplitude, respectively).

Equation (\ref{eq:coeff_ode_multimode_supp}) forms the basis of the model. The advantages of this model are the simultaneous energy conservation and the description of the phase-matching and recoil effects at long interaction lengths, as well as the possibility to construct the exact analytical solution. The model also creates another bridge between free-electron quantum optics and conventional quantum nonlinear optics.

\subsection{Schr\"odinger equation}
We now consider the initial state of the cavity as a single-mode Fock state $\ket{n}$ and the electron initial state $\ket{\Omega_0} \equiv \ket{m=0}$. From Eq.~\ref{eq:coeff_ode_multimode_supp} we can get
\begin{equation}
\begin{aligned}
\frac{\mathrm{d}}{\mathrm{d} L} C_{m} = \ & \tilde{g}_{Qu} e^{-i(k(\Omega_{m})-k(\Omega_{m-1})-\varkappa)L} \sqrt{l+1} C_{m-1} \\ &- \tilde{g}_{Qu}^* e^{i(k(\Omega_{m+1})-k(\Omega_{m})-\varkappa)L} \sqrt{l} C_{m+1}, \\ l = 0,1,2&,..; \ m = 0, \pm 1, \pm 2,..; \ l+m=n.
  \label{eq:coeff_ode_n}
\end{aligned}
\end{equation}
Due to the Fock initial state of the cavity and the perfect entanglement, $\bar{C}$ is now a column and not a matrix.

If the initial state of the cavity is a single-mode vacuum state $\ket{0}$, we get
\begin{equation}
\begin{aligned}
\frac{\mathrm{d}}{\mathrm{d} L} C_{-n} = \ & \tilde{g}_{Qu} e^{-i(k(\Omega_{-n})-k(\Omega_{-n-1})-\varkappa)L} \sqrt{n+1} C_{-n-1} \\ &- \tilde{g}_{Qu}^* e^{i(k(\Omega_{-n+1})-k(\Omega_{-n})-\varkappa)L} \sqrt{n} C_{-n+1}. \ \ (n = 0,1,2..)
  \label{eq:coeff_ode_sup}
\end{aligned}
\end{equation}
Or in the matrix form:
\begin{equation}
\begin{aligned}
&\frac{\mathrm{d}}{\mathrm{d} L} \bar{C}(L) = \bar{\bar{A}} (L)\cdot \bar{C}(L), \ \ \ \bar{C}(L)= \begin{bmatrix}
  C_0(L), C_{-1}(L), C_{-2}(L), \cdots
  \end{bmatrix}^\mathrm{T} 
  \\ &\bar{\bar{A}} (L) =
  \begin{bmatrix}
    0 & \tilde{g}_{Qu}e^{-i\Delta k_{-1} L} & 0 & \cdots \\
    -\tilde{g}_{Qu}^* e^{i\Delta k_{-1} L} & 0 & \tilde{g}_{Qu}e^{-i\Delta k_{-2}L}\sqrt{2} & \cdots \\
    0 & -\tilde{g}_{Qu}^* e^{i\Delta k_{-2}L}\sqrt{2} & 0 & \cdots \\
    \vdots & \vdots & \vdots &\ddots\\
  \end{bmatrix}.
  \label{eq:coeff_ode_matrix}
\end{aligned}
\end{equation}
Here $\Delta k_n = k(\Omega_{-n+1})-k(\Omega_{-n})-\varkappa$, $C_{-n}$ is the probability amplitude of the entangled electron-photon state $\ket{\Omega_0-n\cdot\omega}\otimes\ket{n}\equiv\ket{\Omega_{-n}, n}$ (electron has energy $\hbar (\Omega_0-n\cdot\omega)$, cavity is in the Fock state with $n$ photons). Thus, $|C_{-n}|^2$ describes both the electron spectrum and cavity photon statistics during the interaction. The evolution matrix $\bar{\bar{A}}$ is an anti-Hermitian matrix (so that matrix exponential $e^{\bar{\bar{A}}}$ is unitary) with only -1 and +1 diagonals filled. 

Eq.~\ref{eq:coeff_ode_n} is a system of non-autonomous ordinary differential equations (ODE). We will solve it both numerically and analytically (via the autonomization procedure). But first we start with the approximate approach and make use of the Magnus expansion \cite{magnus_exponential_1954}. In the first order of Magnus expansion the solution of Eq.~\ref{eq:coeff_ode_matrix} is:
\begin{equation}
\bar{C}(L) \approx \exp \left( \int_{0}^L \bar{\bar{A}}(s)\,\mathrm{d}s \right) \cdot \bar{C}(0).
\label{eq:magnus}
\end{equation}
Integration of the matrix $\bar{\bar{A}}$ leads to the terms of the form 
\begin{equation}
\tilde{g}_{\rm Qu}\int_{0}^L e^{-i(k(\Omega_{-n+1})-k(\Omega_{-n})-\varkappa)s} \mathrm{d}s = g_{\rm Qu} \sinc(\Delta k_n L/2)e^{-\Delta k_n L/2},
\end{equation}
where $\Delta k_n = k(\Omega_{-n+1})-k(\Omega_{-n})-\varkappa$. Common phase term doesn't influence the statistics, so now levels have the effective coupling modulated by the mismatch $g_{\mathrm{eff} \ (n)}^{\mathrm{sinc}} = g_{\rm Qu} \sinc(\Delta k_n L/2)$. This approach (noted as sinc-model) can be useful in the low-coupling regime, though giving a slightly tighter recoil cut-off than the exact solution (Fig.~\ref{fig:1photon}c). In the strong-coupling regime the effective phase-matching width is expected to depend on the coupling strength \cite{sharapova_properties_2020} — this behaviour is observed for the exact solution, but not for the sinc-model.

\subsection{Autonomization}
Eq.~\ref{eq:coeff_ode_n} can be solved analytically via the autonomization procedure. We substitute $C_{m}(L)\equiv f_{m}(L) e^{-i\varphi_m L}$, so from Eq.~\ref{eq:coeff_ode_n} 
\begin{equation}
\begin{aligned}
f_{m}' - i\varphi_m f_{m} = \ & \tilde{g}_{Qu} e^{-i(\Delta k_{m-1}+\varphi_{m-1}-\varphi_{m})L} \sqrt{l+1} f_{m-1} \\ &- \tilde{g}_{Qu}^* e^{i(\Delta k_{m}-\varphi_{m+1}+\varphi_{m})L} \sqrt{l} f_{m+1},\\ l = \ 0,1,2..; \ &m = 0, \pm 1, \pm 2,..; \ l+m=n,
  \label{eq:auto1}
\end{aligned}
\end{equation}
where $\Delta k_{m} = k(\Omega_{m+1})-k(\Omega_{m})-\varkappa$. To get rid of the $L$-dependence on the right-hand side we demand $\varphi_{m}-\varphi_{m-1}=\Delta k_{m-1}$ and $\varphi_{m+1}-\varphi_{m}=\Delta k_{m}$. From this we get
\begin{equation*}
\varphi_{m} = 
\begin{cases}
 \varphi_{0}-\sum_{i=-1}^{m}\Delta k_{i} & \text{if $m<0$} \\
 \varphi_{0}+\sum_{i=0}^{m-1}\Delta k_{i} & \text{if $m>0$}.
\end{cases}
\end{equation*}
Since $\varphi_0$ is a global phase, we can take $\varphi_0=0$. As a result, Eq.~\ref{eq:auto1} becomes autonomous
\begin{equation*}
\begin{aligned}
f_{m}' = i\varphi_m f_{m} + \tilde{g}_{Qu}\sqrt{l+1} f_{m-1} - \tilde{g}_{Qu}^*\sqrt{l} f_{m+1} \ \ \longleftrightarrow \ 
\frac{\mathrm{d}}{\mathrm{d} L} \bar{f}(L) = \bar{\bar{S}} \cdot \bar{f}(L),
  \label{eq:auto_ode}
\end{aligned}
\end{equation*}
and we can write the exact analytical solution 
\begin{equation}
\begin{aligned}
\bar{f}(L) = e^{\bar{\bar{S}}\cdot L}\cdot \bar{f}(0)\ \ \Rightarrow \ \ 
\bar{C}(L) = \bar{f}(L) e^{-i\bar{\varphi} L}.
  \label{eq:auto_sol}
\end{aligned}
\end{equation}
The evolution matrix $\bar{\bar{S}}$ is a tridiagonal anti-Hermitian matrix (so the matrix exponential $e^{\bar{\bar{S}}\cdot L}$ is unitary). Note that for the calculation of the electron-photon state probability distribution $|C_{m}|^2 = |f_{m}|^2$, so the phase factor $e^{-i\varphi_m L}$ is not important, thought for the fidelity calculations this factor matters. We can compare the solution after the autonomization with the numerical solution of the initial Eq.~\ref{eq:coeff_ode_matrix} to make sure that they are in the perfect agreement (Fig.~\ref{fig:1photon}c).

If the initial state of the cavity is a vacuum state $\ket{0}$, we can write
\begin{equation}
\begin{gathered}
f_{-n}' = i\varphi_{-n} f_{-n} + \tilde{g}_{Qu}\sqrt{n+1} f_{-n-1} - \tilde{g}_{Qu}^*\sqrt{n} f_{-n+1}, \ \ \ (n = 0,1,2..)
  \\ \bar{\bar{S}}\cdot L =
  \begin{bmatrix}
    i\varphi_0 L =0 & g_{\rm Qu} & 0 & \cdots \\
    -g_{\rm Qu}^*& i\varphi_{-1}L & g_{\rm Qu}\sqrt{2} & \cdots \\
    0 & -g_{\rm Qu}^*\sqrt{2} & i\varphi_{-2}L & \cdots \\
    \vdots & \vdots & \vdots &\ddots\\
  \end{bmatrix}.
  \label{eq:autonomized}
\end{gathered}
\end{equation}

\clearpage
\section*{Quantum simulations}
\section{Lattice Hamiltonian derivation}\label{sec:Supp_lattice}

In this section, we outline how the effective ladder Hamiltonians used in the main text are obtained from the full continuum interaction Hamiltonian of a free electron coupled to the electromagnetic field. Throughout, we work in a length-propagation picture, in which the propagation distance \(L\) (or equivalently \(z\)) along the electron trajectory plays the role of the evolution parameter.

Starting from the Heisenberg equations of motion for the electron annihilation operator,
\begin{equation}
v\,\frac{\mathrm{d}}{\mathrm{d}L}\,\hat{c} = \frac{i}{\hbar}[\hat{H},\hat{c}],
\end{equation}
and using the relativistic Dirac field of the free electron minimally coupled to the quantized electromagnetic field, one obtains the interaction Hamiltonian in the continuum energy representation,
\begin{equation}
\begin{aligned}
\hat{H}_{\mathrm{int}}(L) &= 2\pi i \hbar v 
\int \mathrm{d}\omega\,\mathrm{d}\Omega\,
\Gamma\,
e^{i\left[k(\Omega)-k(\Omega-\omega)-\varkappa(\omega)\right]L}\,
\hat{c}^{\dagger}(\Omega-\omega,L)\hat{c}(\Omega,L)\hat{a}^{\dagger}(\omega,L) \\
&\quad - 2\pi i \hbar v 
\int \mathrm{d}\omega\,\mathrm{d}\Omega\,
\Gamma^{*}\,
e^{-i\left[k(\Omega+\omega)-k(\Omega)-\varkappa(\omega)\right]L}\,
\hat{c}^{\dagger}(\Omega+\omega,L)\hat{c}(\Omega,L)\hat{a}(\omega,L),
\end{aligned}
\label{eq:Hamiltonian_L}
\end{equation}
where \(\hat{c}(\Omega,L)\) and \(\hat{a}(\omega,L)\) annihilate an electron of energy \(\hbar\Omega\) and a photon of energy \(\hbar\omega\), respectively. Here \(k(\Omega)\) is the longitudinal electron wavevector, \(\varkappa(\omega)\) is the photon wavevector of the optical mode, \(\Gamma\) is the microscopic coupling constant, and \(v\) is the electron group velocity. The exponential phase factors encode longitudinal momentum mismatch accumulated during propagation.

To connect this continuum description to the ladder Hamiltonians used in the main text, we discretize the electron and photon spectra by expanding the electron field in a basis of recoil-resolved energy sidebands. Specifically, we introduce ladder states \(\ket{m}\) corresponding to electron energies \(\hbar(\Omega_0-m\omega)\), where \(\Omega_0\) is the initial electron energy and \(\omega\) denotes the relevant photon frequency. Restricting to a discrete set of photon modes and absorbing constant prefactors into a mode-dependent coupling \(\gamma_\omega\), the interaction Hamiltonian takes the discrete form
\begin{equation}
\begin{aligned}
\hat{H}_{\mathrm{int}}(L) &= i\hbar v 
\sum_{\omega,\Omega}
\gamma_{\omega}\,
e^{i\left[k(\Omega)-k(\Omega-\omega)-\varkappa(\omega)\right]L}\,
\hat{c}^{\dagger}_{\Omega-\omega}(L)\hat{c}_{\Omega}(L)\hat{a}^{\dagger}_{\omega}(L) \\
&\quad - i\hbar v 
\sum_{\omega,\Omega}
\gamma_{\omega}^{*}\,
e^{-i\left[k(\Omega+\omega)-k(\Omega)-\varkappa(\omega)\right]L}\,
\hat{c}^{\dagger}_{\Omega+\omega}(L)\hat{c}_{\Omega}(L)\hat{a}_{\omega}(L),
\end{aligned}
\label{eq:Hamiltonian_L_discrete}
\end{equation}
which corresponds to Eq.~(\ref{eq:Hamiltonian_L}) expressed in a recoil-resolved, discrete-energy basis.

In the driven (pumped) regime considered in the quantum simulation sections of the main text, the photon operators \(\hat{a}_\omega\) are replaced by coherent amplitudes associated with externally applied optical tones. This reduces the interaction to an effective single-particle Hamiltonian acting on the electron ladder alone,
\begin{equation}
H(z)
=
\sum_{m,j}
\left[
J_j\,e^{i\left(k_{m+1}-k_m-\varkappa_j\right)z}\,
\ket{m{+}1}\!\bra{m}
+
\mathrm{h.c.}
\right],
\end{equation}
where \(m\) labels ladder states differing by one photon recoil, \(\varkappa_j\) and \(J_j\) denote the wavevector and complex amplitude of the \(j\)th optical tone, and \(z=L\) has been identified with the propagation coordinate. This is the non-autonomous Hamiltonian used throughout the main text.

Finally, in the high-recoil regime, rapidly oscillating off-resonant terms average to zero, yielding an effective autonomous tight-binding Hamiltonian with programmable nearest-neighbor couplings. The sequence of approximations outlined above establishes a direct and controlled connection between the full relativistic electron–photon interaction and the simplified ladder Hamiltonians employed for quantum simulation and quantum information processing.

\subsection{Reduction to an effective lattice Hamiltonian}

Equation~(\ref{eq:H_nonauto}) describes a driven, non-autonomous evolution of the electron ladder in the propagation coordinate \(z\), with coupling between all neighboring ladder states mediated by globally applied optical tones. The phase factors
\(
\exp[i(k_{m+1}-k_m-\varkappa_j)z]
\)
encode the accumulated longitudinal momentum mismatch between the electron recoil transition \(\ket{m}\rightarrow\ket{m+1}\) and the applied optical field.

To obtain a simplified description suitable for quantum simulation and Hamiltonian engineering, we consider the high-recoil or strong-mismatch regime, in which
\begin{equation}
\big|k_{m+1}-k_m-\varkappa_j\big|\,L \gg 1
\qquad
\text{for all off-resonant links},
\end{equation}
over the interaction length \(L\). In this limit, the rapidly oscillating off-resonant terms in Eq.~(\ref{eq:H_nonauto}) average to zero upon propagation, while only near-resonant contributions survive. Physically, this corresponds to momentum-selective addressing of ladder transitions via recoil, without requiring local control in real space.

Under these conditions, the dynamics reduce to an effective autonomous Hamiltonian acting on the discrete ladder alone,
\begin{equation}
H_{\mathrm{eff}}
=
\sum_m
J_m
\left(
\ket{m{+}1}\!\bra{m}
+
\ket{m}\!\bra{m{+}1}
\right),
\label{eq:H_eff_lattice}
\end{equation}
where the effective nearest-neighbor couplings \(J_m\) are determined by the amplitudes and phases of the applied optical tones chosen to satisfy the corresponding phase-matching conditions. Equation~(\ref{eq:H_eff_lattice}) is a tight-binding Hamiltonian on a one-dimensional lattice indexed by the recoil ladder number \(m\), with fully programmable couplings.

The mapping in Eq.~(\ref{eq:H_eff_lattice}) provides the basis for all quantum simulation protocols discussed in the main text. By engineering the spatial profile of the couplings \(J_m\), one can implement a wide class of lattice Hamiltonians, including uniform chains, finite-dimensional qudit registers, and position-dependent coupling landscapes that emulate effective curved-spacetime geometries. Importantly, because the control parameters are optical amplitudes and phases, the couplings \(J_m\) can be reconfigured dynamically and on ultrafast timescales.

Finally, we note that independent control over the magnitudes and phases of the couplings \(J_m\) implies full controllability on any finite truncation of the ladder. As shown in the Supplementary Information, the resulting set of generators closes under commutation and yields universal control both in the qudit setting and in the Lloyd–Braunstein framework for universal quantum simulation of continuous-variable systems \cite{lloyd1999quantum}. This establishes a direct and systematic connection between the first-principles electron–photon interaction and the effective lattice Hamiltonians employed throughout the main text.

\section{Quantum simulation of Hawking radiation and curved spacetime with a recoiled free electron}\label{sec:Supp_black_holes}

The recoiled free-electron platform provides a natural setting for quantum simulation of effective curved spacetimes and horizon physics. In particular, the electron energy ladder induced by photon recoil can be interpreted as a synthetic spatial dimension, while controlled photon–electron interactions generate programmable hopping amplitudes along this dimension. This enables a direct analogue of quantum walks in curved spacetime, closely related to recent superconducting-circuit realizations of analogue black holes, but implemented here within a single propagating quantum particle with full and ultrafast optical control.

\begin{figure}[ht]
\centering
\includegraphics[width=1\linewidth]{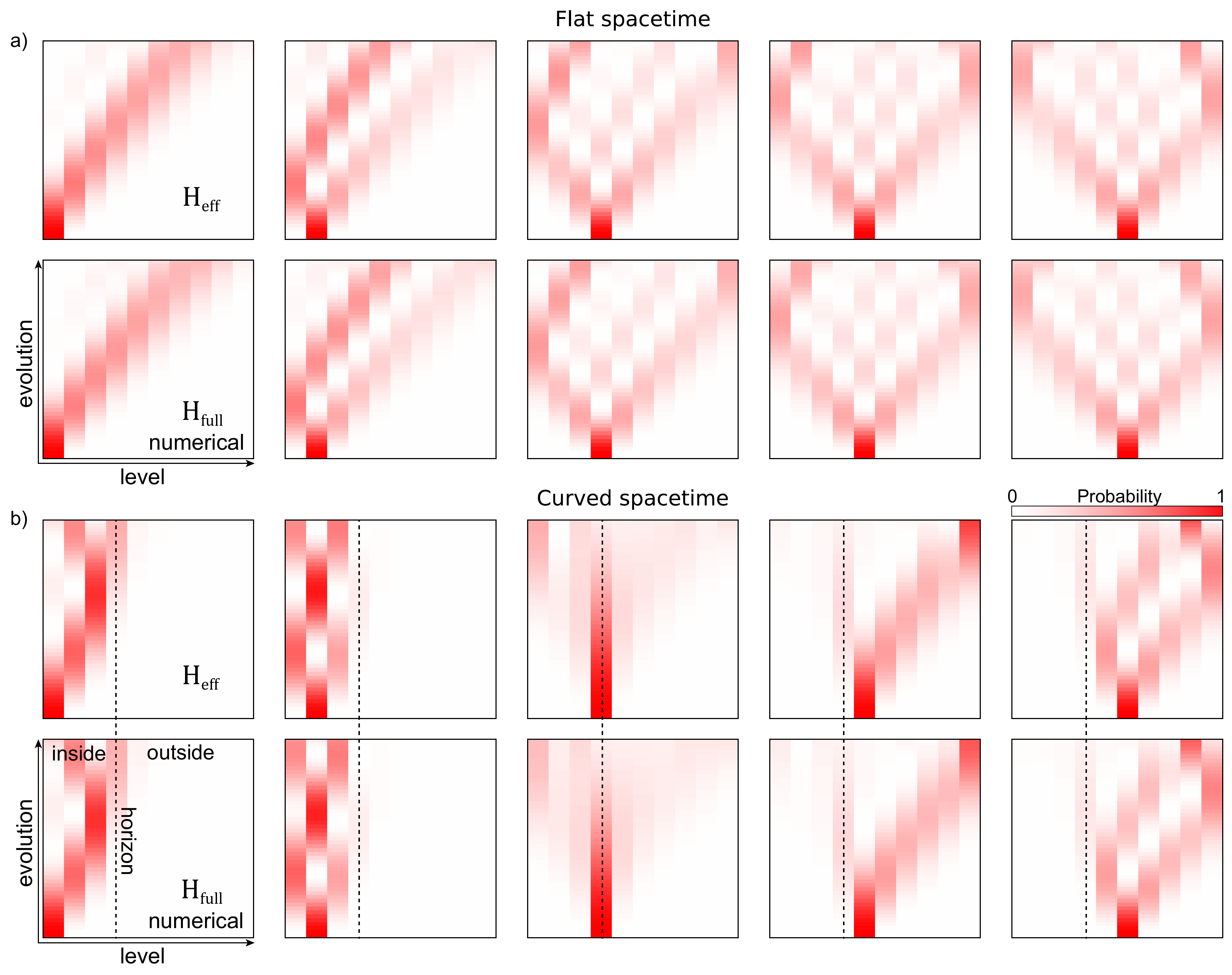}
\caption{\label{fig:spacetime} Quantum simulation of Hawking radiation and curved spacetime with a recoiled free electron. a) Flat spacetime: $H_{\rm eff}$ analytical solution (top row) and $H_{\rm full}$ numerical solution (bottom row). b) Curved spacetime: $H_{\rm eff}$ analytical solution (top row) and $H_{\rm full}$ numerical solution (bottom row). Event horizon is placed at $m=3$. Here a single interaction zone with $L=10^{-3}~\rm m$, and an electron with $1~\rm keV$ is used.}
\end{figure}

\subsection{Effective metric, Painlev\'e--Gullstrand black holes, and the ladder Hamiltonian}

Analogue models of black-hole horizons are most naturally formulated in terms of effective spacetime metrics experienced by propagating excitations. In one spatial dimension, a particularly convenient representation is the Painlev\'e--Gullstrand (PG) form of the black-hole metric,
\begin{equation}
ds^2
=
-\left[c^2 - v(x)^2\right]\,dt^2
-2\,v(x)\,dt\,dx
+dx^2,
\label{eq:PG_metric}
\end{equation}
where $c$ is the characteristic propagation speed of excitations and $v(x)$ is a position-dependent background flow velocity. A black-hole horizon occurs at the location $x=x_h$ where $|v(x_h)|=c$. Importantly, the PG metric is regular at the horizon and admits a clear physical interpretation in terms of a flowing medium, making it the standard description for analogue black holes in condensed-matter, optical, and circuit-based platforms.

Near the horizon, the flow velocity can be linearized as
\begin{equation}
v(x) \simeq c - \kappa (x-x_h),
\end{equation}
where $\kappa$ is the surface gravity. In this limit, the metric reduces locally to that of Rindler spacetime, which describes a uniformly accelerated frame and captures the universal near-horizon physics responsible for Hawking radiation. On a lattice, this Rindler region is typically regularized by a smooth interpolation, most commonly implemented using a hyperbolic-tangent profile.

In lattice quantum-walk realizations, the PG metric maps onto a position-dependent hopping Hamiltonian. In the single-particle sector, the effective Hamiltonian takes the form
\begin{equation}
H_{\mathrm{PG}}
=
\sum_m
J_m
\left(
\ket{m{+}1}\!\bra{m}
+
\ket{m}\!\bra{m{+}1}
\right),
\label{eq:H_PG_lattice}
\end{equation}
where the discrete index $m$ labels spatial sites and the hopping amplitudes $J_m$ encode the background flow velocity. A smooth Rindler horizon is realized by choosing
\begin{equation}
J_m
=
\frac{\beta}{4\eta_d}
\tanh\!\left[(m-m_h+\tfrac12)\eta_d\right],
\label{eq:Rindler_profile}
\end{equation}
with $m_h$ the horizon position, $\eta_d$ controlling the sharpness of the transition, and $\beta$ setting the overall coupling scale. This Hamiltonian supports asymmetric mode propagation, horizon-induced mode conversion, and Hawking-like radiation.

In the recoiled free-electron platform, the role of the spatial coordinate is played by the electron’s energy ladder index $m$, corresponding to photon recoil sidebands. The electron propagates in real space, so the natural evolution parameter is the propagation distance $z$. The electron state $\ket{\psi(z)}$ therefore obeys a length-propagation Schr\"odinger equation,
\begin{equation}
i\,\frac{d}{dz}\ket{\psi(z)} = H(z)\ket{\psi(z)}.
\end{equation}
Before any rotating-frame transformation, interaction with optical fields leads to the explicitly non-autonomous Hamiltonian
\begin{equation}
H(z)
=
\sum_{m,j}
\left[
J_j
\,e^{i\left(k_{m+1}-k_m-\varkappa_j\right)z}\,
\ket{m{+}1}\!\bra{m}
+
\mathrm{h.c.}
\right],
\label{eq:H_nonauto_PG}
\end{equation}
where $k_m$ is the electron wavevector in ladder state $\ket{m}$, $\varkappa_j$ is the wavevector of the $j$th applied optical tone, and $J_j$ its coupling amplitude. Each tone couples globally to all ladder transitions, while momentum mismatch $k_{m+1}-k_m-\varkappa_j$ enforces selectivity through rapidly oscillating phases.

In the high-recoil limit, where $|k_{m+1}-k_m-\varkappa_j|L\gg1$ for all off-resonant links over a propagation length $L$, the oscillatory terms average to zero. The dynamics then reduce to an effective autonomous Hamiltonian of the form
\begin{equation}
H_{\mathrm{eff}}
=
\sum_m
J_m
\left(
\ket{m{+}1}\!\bra{m}
+
\ket{m}\!\bra{m{+}1}
\right),
\end{equation}
which is identical to the lattice PG Hamiltonian in Eq.~\eqref{eq:H_PG_lattice}. By programming the amplitudes of the applied optical tones, one directly engineers the effective coupling profile $J_m$, thereby simulating a one-dimensional black hole in the recoiled electron’s energy ladder.

\subsection{Comparison to superconducting-circuit black-hole simulators}

The present approach is closely related in spirit to recent quantum simulations of black-hole horizons implemented in superconducting circuits, where a chain of tunable qubits or resonators realizes a position-dependent hopping Hamiltonian with an effective horizon \cite{shi2023quantum}. In those systems, curved spacetime is engineered by dynamically programming inter-site couplings and onsite energies, and Hawking radiation manifests as correlated excitations propagating away from the horizon on opposite sides of the chain. The recoiled free-electron platform realizes an analogous lattice Hamiltonian, but within a fundamentally different physical setting.

A key distinction is that the electron ladder constitutes a {synthetic spatial dimension embedded in a single propagating quantum particle}, rather than a collection of spatially separated qubits. The effective sites $\ket{m}$ correspond to energy sidebands of the same electron, and the hopping amplitudes are programmed optically through momentum-selective photon recoil. As a result, the electron naturally propagates through the entire simulated spacetime in a single pass, without the need for local control wiring, cryogenic infrastructure, or time-dependent tuning of many individual elements. In addition, the free-electron system supports a much larger number of accessible ladder states, enabling simulations in higher-dimensional Hilbert spaces with long intrinsic coherence lengths set by free-space propagation rather than material relaxation times.

From a Hamiltonian perspective, superconducting-circuit simulators typically implement an explicitly time-dependent control sequence to engineer the desired coupling profile, whereas in the recoiled-electron platform the curved-spacetime Hamiltonian emerges either directly from the non-autonomous propagation-picture Hamiltonian, or, in the high-recoil limit, as an effective autonomous Hamiltonian. This provides a complementary route to analogue gravity, in which spatial variation and momentum mismatch replace temporal modulation as the primary control resource.

\subsection{Hawking pair correlations and entanglement}

Beyond single-particle transport, the recoiled-electron platform naturally enables the study of Hawking-like pair correlations and entanglement. In analogue-gravity systems, Hawking radiation arises from mode conversion at the horizon, leading to correlated pairs of excitations emitted on opposite sides of the horizon. In superconducting-circuit realizations, this physics is revealed through correlations between qubit or resonator excitations inside and outside the effective black hole.

In the recoiled-electron system, analogous correlations appear in the joint electron–photon degrees of freedom and, more generally, in correlations between different ladder modes. The non-autonomous Hamiltonian explicitly mixes ladder states with momentum-dependent phases, enabling the conversion of incoming modes into superpositions of outgoing ladder excitations. When multiple optical modes or vacuum fluctuations are included, this process produces correlated excitations analogous to Hawking pairs, with one component propagating toward increasing $m$ (lower electron kinetic energy) and its partner propagating toward decreasing $m$.

Importantly, because the ladder states belong to a single electron, these correlations are intrinsically phase coherent and can be accessed interferometrically. By preparing superpositions of ladder states or by measuring joint electron–photon observables, one can directly probe entanglement generated at the effective horizon. In this sense, the recoiled-electron platform provides a unified setting in which Hawking-like radiation, mode entanglement, and curved-spacetime quantum walks can be investigated within a single, highly controllable quantum system. This opens a route to studying horizon-induced entanglement and information flow in analogue gravity using tools from quantum optics and quantum information, complementing existing superconducting and atomic implementations.

\section{Universality}\label{sec:Supp_universality}
\subsection{Finite-dimensional controllability}

On any finite truncation $\mathcal{H}_N=\mathrm{span}\{\ket{0},\ldots,\ket{N}\}$:

\begin{itemize}
\item Each complex $J_m$ generates $\mathfrak{su}(2)$ control on the subspace $\{\ket{m},\ket{m{+}1}\}$.
\item By concatenating adjacent controls, arbitrary Givens rotations between any pair $\ket{m},\ket{n}$ can be synthesized.
\item Virtual $Z$ rotations (diagonal phases) are absorbed by redefining the phases of subsequent drives.
\end{itemize}

Therefore, the reachable Lie algebra is $\mathfrak{su}(N{+}1)$, implying full controllability on $\mathcal{H}_N$.

\subsection{Continuous-variable universality}

Because the system is controllable on \emph{every} finite truncation, it can approximate any unitary generated by polynomial Hamiltonians of $a,a^\dagger$.
The elementary swap operations are number-selective and therefore intrinsically non-Gaussian.

Hence, in the limit $N\to\infty$, the platform is universal for continuous-variable quantum computation in the Lloyd–Braunstein sense \cite{lloyd1999quantum}, without requiring an explicit cubic phase gate \cite{eriksson2024universal}.

\section{Gates Demonstrated}\label{sec:Supp_gates}
We quantify the performance of all implemented operations using an error metric that accounts for both leakage and unitary imperfections. Specifically, we define the gate error rate as
\begin{equation}
\varepsilon \;\equiv\; 1 - p_{\mathrm{stay}}\,F_{\mathrm{unit}},
\end{equation}
where \(p_{\mathrm{stay}}\) is the probability that the system remains within the intended computational subspace after the operation, and \(F_{\mathrm{unit}}\) is the unitary gate fidelity conditioned on staying in that subspace. The stay probability \(p_{\mathrm{stay}}\) captures population leakage into ladder levels outside the encoded qudit or qubit subspace.

The conditional unitary fidelity is defined as
\begin{equation}
F_{\mathrm{unit}}
=
\frac{1}{d^2}\,
\left|\mathrm{Tr}\!\left(U_{\mathrm{target}}^{\dagger}\,U_{\mathrm{eff}}\right)\right|^2,
\end{equation}
where \(d\) is the dimension of the computational subspace, \(U_{\mathrm{target}}\) is the ideal target unitary acting on that subspace, and \(U_{\mathrm{eff}}\) is the effective unitary obtained by projecting the full evolution operator onto the computational subspace and renormalizing to remove leakage. In this formulation, \(F_{\mathrm{unit}}\) measures the coherence of the implemented operation within the retained subspace, while \(p_{\mathrm{stay}}\) quantifies the probability of remaining in that subspace. 

As expected, the dominant error mechanism in these gates arises from population leakage outside the encoded subspace. In the recoiled-electron ladder, this leakage decreases quadratically with the interaction length, leading to an overall error scaling proportional to \(1/L^2\). 

\subsection{Adjacent SWAP}

A $\pi$-pulse on a single link implements
\begin{equation}
\ket{m}\leftrightarrow\ket{m{+}1},
\end{equation}
with tunable complex phase.
This establishes coherent nearest-neighbor control and is the primitive Givens rotation.

\subsection{Non-adjacent SWAP}

Concatenating adjacent swaps yields effective
\begin{equation}
\ket{m}\leftrightarrow\ket{m{+}k},
\end{equation}
demonstrating compiled long-range control and scalability of the ladder.

\subsection{Virtual $Z$ phases}

Diagonal phases accumulated during evolution are tracked classically and absorbed into subsequent drives by adjusting $\arg(J_m)$.
Thus, explicit physical $Z$ gates are unnecessary.

\subsection{Encoded CZ gate}

Encoding two logical qubits in the subspace
\begin{equation}
\{\ket{3},\ket{4},\ket{5},\ket{6}\}
\;\leftrightarrow\;
\{\ket{00},\ket{01},\ket{10},\ket{11}\},
\end{equation}
we synthesize a controlled-$Z$ gate using $2\pi$ rotation on ancilla level and using $7$–$9$ interaction zones via optimal control.

The gate is characterized by the conditional phase invariant
\begin{equation}
\gamma
=
\theta_{00}-\theta_{01}-\theta_{10}+\theta_{11},
\end{equation}
which is driven to $\gamma=\pi$.
The resulting operation achieves high fidelity up to local $Z$ rotations with negligible leakage.

\subsection{Role of Optimal Control}
Optimal control naturally exploits interference between zones to:
\begin{itemize}
\item accumulate the desired conditional phase,
\item suppress residual mixing within the code space,
\item minimize leakage.
\end{itemize}

This approach is standard in modern quantum hardware and is well matched to the zone-based control model. The effective length of each interaction zone can be controlled by the duration and timing of drive optical pulses.

\subsection{Conclusion}

Starting from a physically motivated non-autonomous Hamiltonian with momentum mismatch, we obtain an autonomous ladder control Hamiltonian with full complex $J_m(t)$ control.
This enables universal control on finite truncations and continuous-variable universality.
Demonstrating adjacent and non-adjacent swaps, virtual $Z$ control, and an encoded CZ gate suffices to establish both single-mode universality and entangling capability.

\clearpage

\clearpage
\section*{Quantum optics}
\section{Electron--photon states}\label{sec:Supp_elphot_states}
\subsection{Table of electron--photon states}
\begin{figure*}[ht]
\centering
\includegraphics[width=0.65\linewidth]{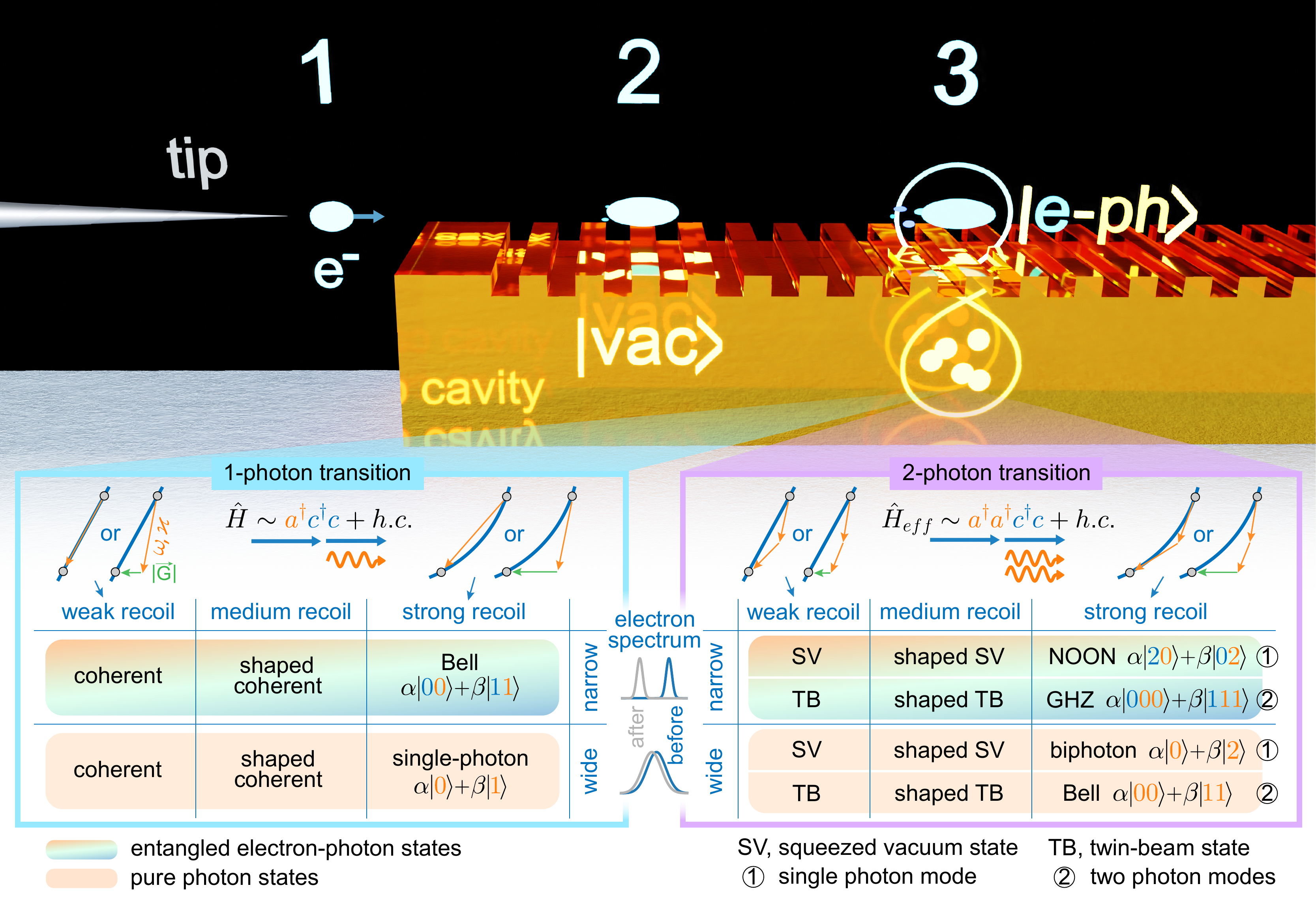}
\caption{\label{fig:abstract} Schematic summary of electron--photon state generation.
{(1)} Free-electron emission (e.g., from a metallic nanotip).
{(2)} Interaction with the evanescent vacuum field of an optical cavity, \(\ket{\mathrm{vac}}\).
{(3)} Formation of an entangled electron--photon state, \(\ket{\mathrm{e\!-\!ph}}\).
Depending on the interaction order (one- or two-photon), recoil strength, and electron spectral width, a variety of quantum states can be generated. The table summarizes the role of electron dispersion (linear: weak recoil; nonlinear: strong recoil) and electron spectral width (overlapping versus distinguishable spectra before and after interaction).
}
\end{figure*}

In general, the generated electron--photon states are entangled, but there are several options where pure photon (and electron) states can be obtained. The first option is to use an initial electron with a broad spectrum, such that the electron spectrum side-bands (connected with emission/absorption of photons) overlap (Fig.~\ref{fig:abstract}). Then the electron states before and after the emission are indistinguishable \cite{huang_electron-photon_2023}, and we can write $\ket{\psi} \approx \ket{\Omega} \otimes (c_0 \ket{0}+c_1 \ket{1}+...)$, so the photon state is pure. The second option is the generation of the deterministic product state $\ket{\psi}=\ket{-1,1}=\ket{-1}\otimes\ket{1}$, in which the states of the photon and electron are by definition pure.

\subsection{Recoil parameter}
\begin{figure}[ht]
\centering
\includegraphics[width=1\linewidth]{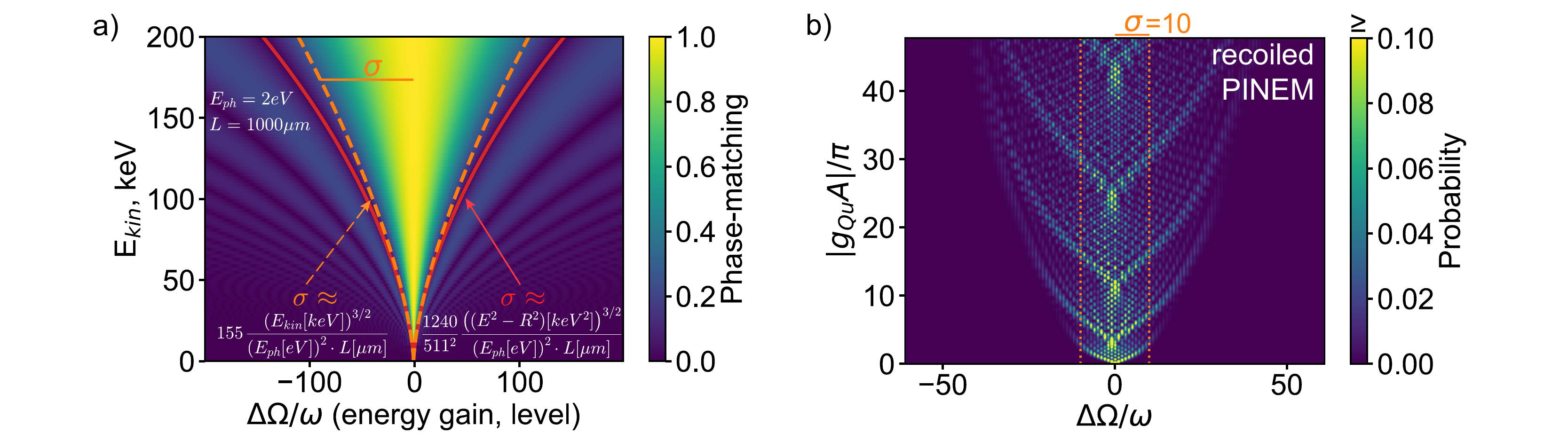}
\caption{\label{fig:phase-matching}a)~Exact calculation of $\sinc(\Delta k_{m}L)$ (map) and estimations of the first zero of the sinc-function with formulas (\ref{eq:sigma_full}, red) and (\ref{eq:sigma_sup}, orange), $E_{\rm ph} = 2~eV$, $L = 1000~\mu m$. b)~PINEM spectra with the recoiled electrons and $\sigma=10$, showing revivals. Parameter $\sigma$ predicts the spectrum boundaries at low coupling strengths. Position (in coupling strength) of the first revival can be estimated from the fitting as $g^{\text{T}_1} \approx 0.84 \sigma + 1.66$, and of the second revival as $g^{\text{T}_2} \approx 3.28 \sigma + 1.95$, where $g^{\text{T}_i} = (|g_{\rm Qu}A|)^{\text{T}_i}$.}
\end{figure}

Sinc-model can provide an important quantitative parameter for assessing the strength of the recoil effect in a given interaction. First, we find the first zero of the sinc function: $\Delta k_{\sigma} L/2 = \pi$. For the interaction with optical photons we can write $\Delta k_{\sigma} \approx \frac{\omega}{c}\frac{1}{\beta_{\sigma}}-\varkappa= \frac{\omega}{c}(\frac{1}{\beta_{\sigma}}-\frac{1}{\beta_1})$, where $\beta_{\sigma}=v^{electron}_{gr \ (\sigma)}/c$ is the normalized electron group velocity after the emission of the $\sigma^{th}$ photon, and we suppose that the emission of the first photon is perfectly phase-matched. Assuming that the total change in the electron energy after the emission of $\sigma$ photons $\delta = \sigma \cdot \hbar \omega$ is much smaller than the initial total electron energy $E = E_{\rm kin}+E_0$ (here $E_{\rm kin}$ is the kinetic energy, $E_0$ is the rest energy of the electron), we can write $\frac{1}{\beta(E-\Delta)}-\frac{1}{\beta(E)} \approx \frac{E_0^2 \Delta}{(E^2-E_0^2)^{3/2}}$. Finally, from $\Delta k_{\sigma} L/2 \approx \frac{\omega L}{2c}\frac{E_0^2 \Delta}{(E^2-E_0^2)^{3/2}} = \pi$ we get
\begin{equation}
\sigma \approx \frac{1240}{511^2} \frac{\left((E^2-E_0^2)[\mathrm{keV}^2]\right)^{3/2}}{\left(E_{\rm ph}[\mathrm{eV}]\right)^2 \cdot L[\mu \mathrm{m}]},
\label{eq:sigma_full}
\end{equation}
which is correct for the wide range of electron energies in SEMs and TEMs (Extended Data Fig.~\ref{fig:phase-matching}a).

Considering electrons with $E_{\rm kin} < 150 \ \mathrm{keV}$ we can get the easier formula:
\begin{equation}
\sigma \approx 155 \frac{\left(E_{\rm kin}[\mathrm{keV}]\right)^{3/2}}{\left(E_{\rm ph}[\mathrm{eV}]\right)^2 \cdot L[\mu \mathrm{m}]}.
\label{eq:sigma_sup}
\end{equation}
Parameter $\sigma$ describes the phase-matching width in the sinc-model: the maximum photon number the electron can emit before the recoil is high enough to completely suppress the phase-matching. Though in the exact solution this border is not sharp and depends on the coupling strength $g_{\rm Qu}$, parameter $\sigma$ gives a good estimation of the recoil effect in a given electron-photon interaction, showing approximately how many photons electron should emit to start experiencing recoil-induced mismatch. From the N-level-system point of view we can approximate the effective number of levels in the system (number of side-bands + zero-loss peak) as 
\begin{equation*}
N_{\mathrm{eff}} = 
\begin{cases}
 \sigma + 1 & \text{if $\sigma \geq 1$} \\
 2 & \text{if $\sigma < 1$},
 \label{eq:neff_sup}
\end{cases}
\end{equation*}
so the transition from the infinite ladder to the two-level system (TLS) can be observed (Fig.~\ref{fig:1photon}a) for different variations of only 3 parameters of the interaction ($E_{\rm kin}, E_{\rm ph}$ and $L$).

\begin{figure}[ht]
\centering
\includegraphics[width=1\linewidth]{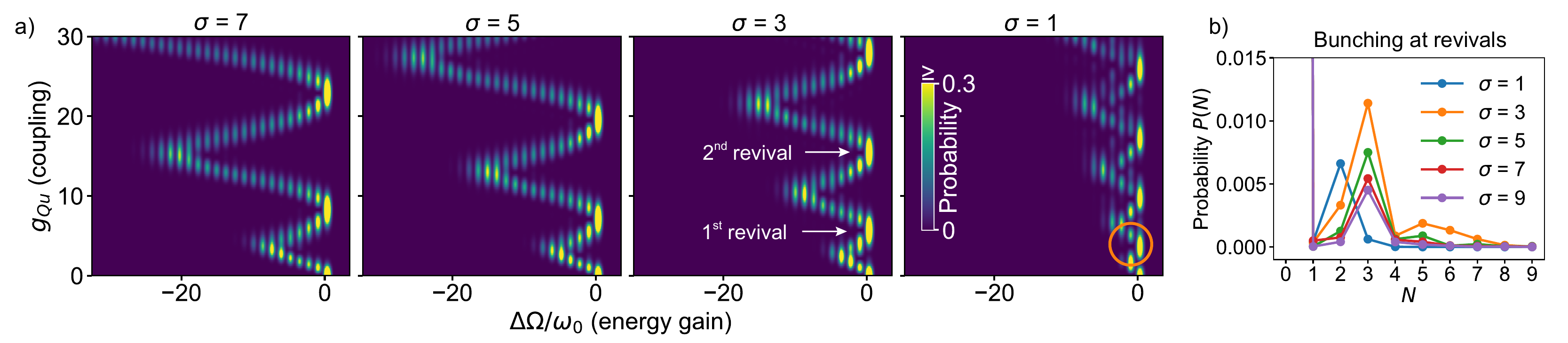}
\caption{\label{fig:superbunching}a)~Revivals of the electron spectrum in the strong coupling regime at different $\sigma$. b)~Generation of k-photon states with high $g^{(2)}$ and low mean photon number $\langle \hat{N} \rangle \ll 1$ at revivals.}
\end{figure}

\begin{figure}[ht]
\centering
\includegraphics[width=1\linewidth]{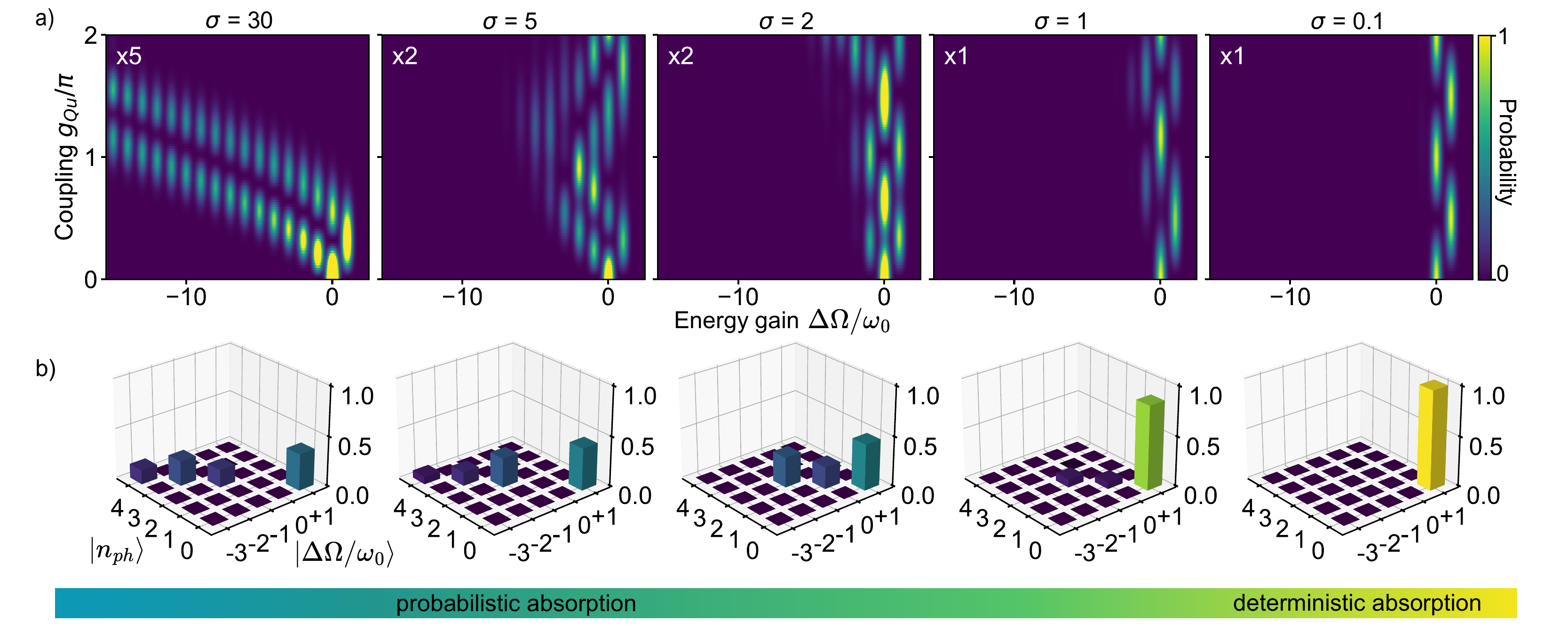}
\caption{\label{fig:absorption} a)~Electron energy spectra from coupling strength $g_{\rm Qu}$ for different phase-matching widths $\sigma$ and cavity in the initial state $\ket{1}$, showing the transition from probabilistic to deterministic single-photon absorption. b)~Probability distribution of electron-photon wavefunction in (a) for states with the maximum single-photon absorption probability.}
\end{figure}

\clearpage
\subsection{Two-photon Hamiltonian}

\begin{figure}[htbp]
\centering
\includegraphics[width=0.75\linewidth]{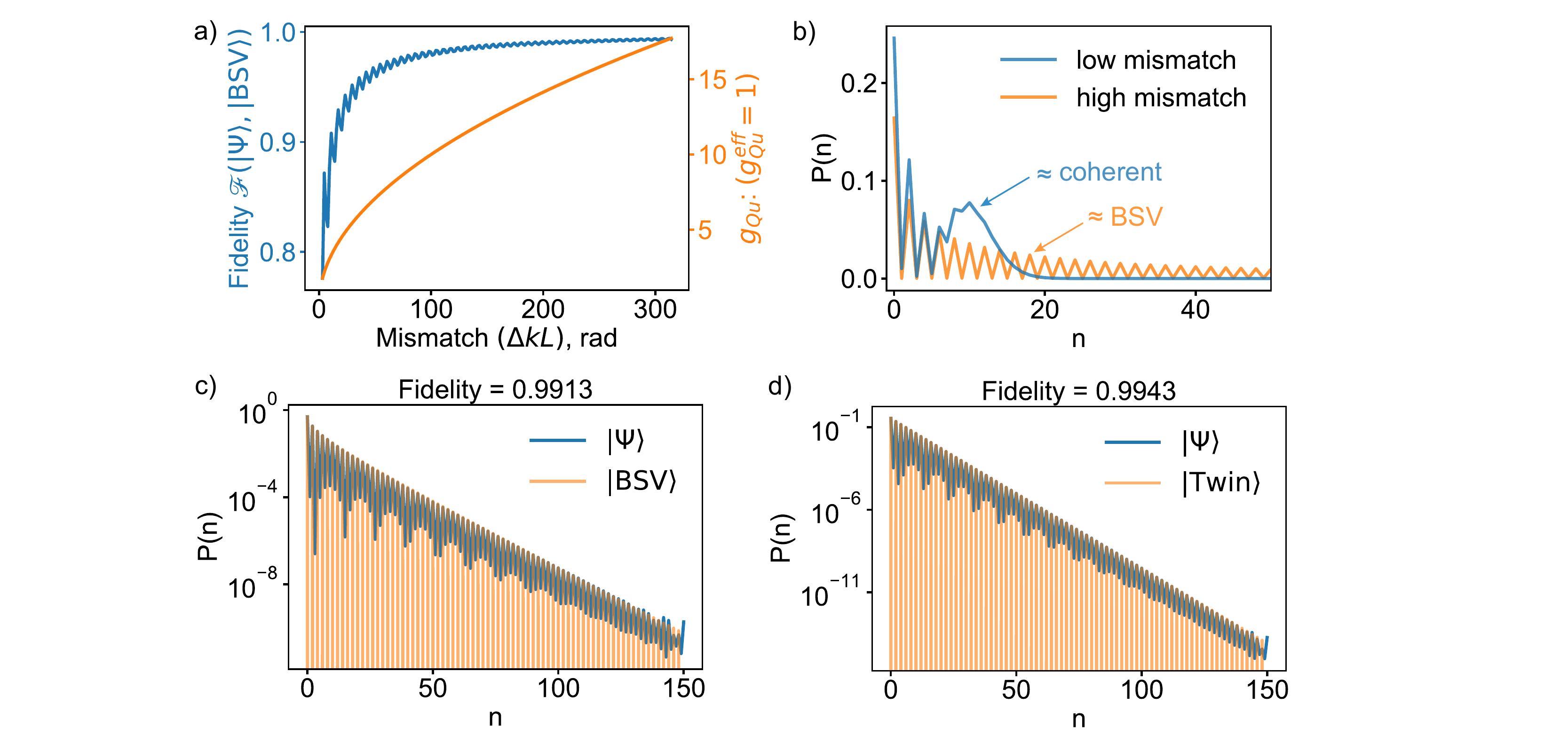}
\caption{\label{fig:fidelity_bsv}a)~Fidelity between the generated state and BSV (blue) along with one-photon $g_{\rm Qu}$ required to achieve two-photon $g_{\rm Qu}^{\mathrm{eff}}=1$ from the one-photon mismatch. Higher fidelity requires higher mismatch and, thus, higher $g_{\rm Qu}$ for the same mean photon number. b)~$P(n)$ for low and high mismatch, showing the transition from coherent-like state (blue) to the BSV-like state (orange). c-d) $P(n)$ for generated BSV and twin-beam states (along with the exact states). Note that twin beam decays uniformly exponentially, while BSV has an accelerated decay at small $n$. The electron-photon squeezed vacuum generation does not require the recoiled electron regime (but requires QPM) and can be achieved in TEMs as well. Twin-beam generation can be achieved also with direct phase-matching.}
\end{figure}

Due to the analogy to the Raman process in the three-level system, we can introduce effective two-photon Hamiltonian $\hat{H}_{\mathrm{eff}}\sim g_{\rm Qu}^{\mathrm{eff}} \cdot \hat{a}^{\dagger}\hat{a}^{\dagger}\hat{c}^{\dagger}\hat{c} + h.c.$ with the coupling strength $g_{\rm Qu}^{\mathrm{eff}}$ for the two-photon process via the adiabatic elimination of the mismatched odd states (taking $\frac{\mathrm{d}}{\mathrm{d} L} C_{-(2m+1)} = 0$, $m \in \mathbb{Z}$). Coupling coefficient between the states $\ket{-m,m}$ and $\ket{-(m+2),m+2}$ after the adiabatic elimination is then $\gamma_m = \frac{\tilde{g}_{Qu}^2}{i\varphi_{-(m+1)}}\cdot \sqrt{m+1}\sqrt{m+2}$, so we can define the effective two-photon coupling strength as
\begin{equation}
\begin{aligned}
g_{\rm Qu}^{\mathrm{eff}} \text{(2-photon)} \equiv \frac{g_{\rm Qu}^2}{-\varphi_{-1}L},
  \label{eq:g_eff}
\end{aligned}
\end{equation}
where $\varphi_{-1}L = -\Delta k_{m}L = -(k(\Omega_{0})-k(\Omega_{-1})-\varkappa)L$ is the one-photon transition mismatch, and $g_{\rm Qu}$ is the one-photon transition coupling. This approach is in agreement with the Raman process notation, where the effective Rabi frequency is $\Omega_{\mathrm{eff}}=\frac{\Omega_{1}\Omega_{2}}{2\Delta}$, and $\Delta$ is the detuning from the intermediate level \cite{scully_quantum_1997, levine_high-fidelity_2018}. The values of $g_{\rm Qu}^{\mathrm{eff}} \text{(2-photon)}$ are also shown in Fig.~\ref{fig:bsv}a. Extended Data Fig.~\ref{fig:fidelity_bsv} demonstrates the influence of the one-photon mismatch on the fidelity between the generated state and perfect BSV, and on the coupling $g_{\rm Qu}^{\mathrm{eff}}$. One-photon mismatch poses the trade-off between state fidelity and $g_{\rm Qu}$ required to achieve the same $g_{\rm Qu}^{\mathrm{eff}}$ (and the same mean photon number). 

The electron-photon single-mode squeezed vacuum state and twin-beam state can be be written in the following form \cite{scully_quantum_1997, lvovsky_squeezed_2015}
\begin{equation}
\begin{gathered}
\ket{{\text{SV}}^{\text{e-ph}}} \equiv \sum_{\Omega, m} C_{\Omega, m} \ket{\Omega, m} = \frac{1}{\sqrt{\cosh r}} \sum_{n=0}^{\infty}\left(-e^{i \varphi} \tanh r\right)^n \frac{\sqrt{(2 n) !}}{2^n n !} \ \ket{-2n, 2n}, \\
\ket{{\text{Twin}}^{\text{e-ph}}} \equiv \sum_{\Omega, m_1, m_2} C_{\Omega, m_1, m_2} \ket{\Omega, m_1, m_2} =\frac{1}{\cosh r} \sum_{n=0}^{\infty}\left(-e^{i \varphi} \tanh r\right)^n \ket{-2n,n,n},
  \label{eq:bsv and twin}
\end{gathered}
\end{equation}
where $r$ and $\varphi$ depend on $g_{\rm Qu}^{\mathrm{eff}}$. These states can be generated with fidelities $\mathscr{F}>99\%$ (see Extended Data Fig.~\ref{fig:fidelity_bsv}). 

\begin{figure}[ht]
\centering
\includegraphics[width=0.9\linewidth]{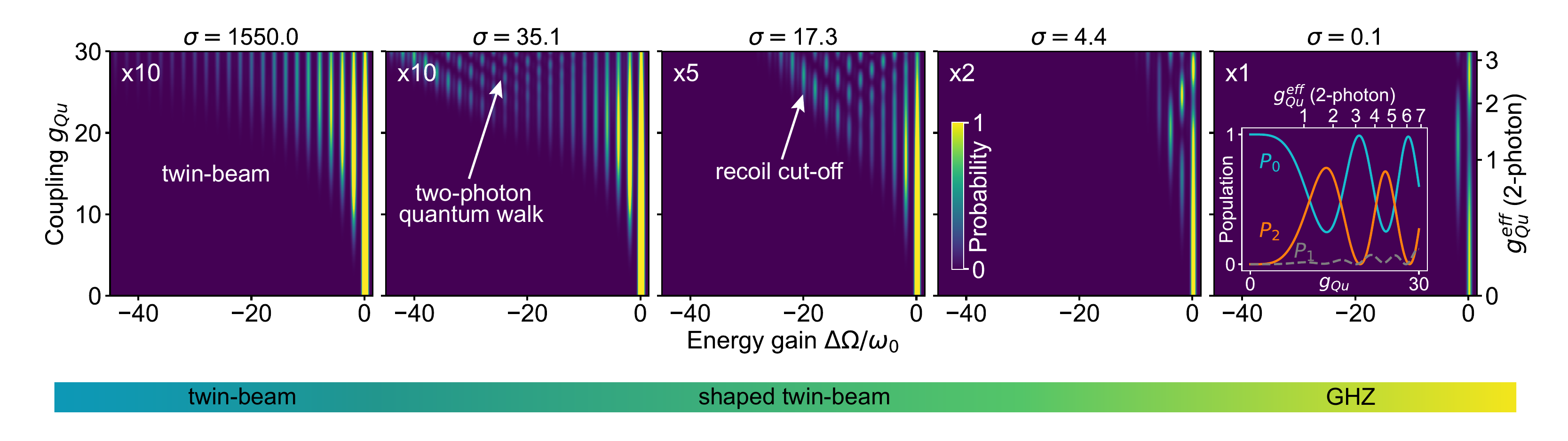}
\caption{\label{fig:twin}Electron energy spectra from coupling strength $g_{\rm Qu}$ for different phase-matching widths $\sigma$, showing the transition from electron-photon twin-beam state to shaped twin-beam state and GHZ states (emission of two photons is phase-matched). The inset (right) shows the two-photon Rabi oscillations for lower one-photon mismatch. }
\end{figure}

\begin{figure}[ht]
\centering
\includegraphics[width=0.9\linewidth]{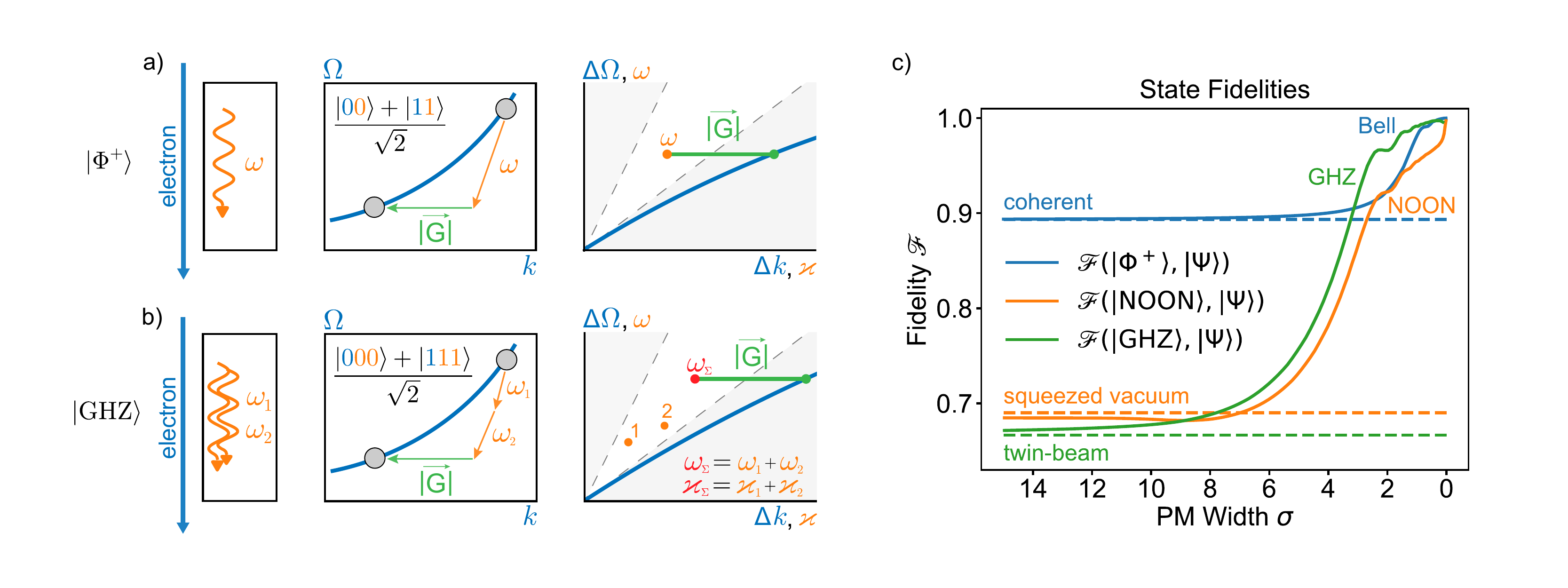}
\caption{\label{fig:GHZ} Electron-photon states at high recoil. a-b) Schemes for the generation of electron-photon Bell (a), GHZ and NOON (b) states, shown here with the quasi-phase-matching (green lines labeled $|\vec{G}|$). If $\omega_1 = \omega_2$ and $\varkappa_1 = \varkappa_2$, then $\ket{\text{GHZ}} \rightarrow \ket{\text{NOON}} (N=2)$. c) Electron-photon Bell, NOON and GHZ states fidelities from the phase-matching width $\sigma$. For high $\sigma$ these fidelities correspond to the fidelities of Bell, NOON and GHZ states with coherent, squeezed vacuum and twin-beam states, respectively (dashed lines). The photon frequencies were the same for all the three states. It can be seen that twin-beam state generation does not require quasi-phase matching, but the squeezed vacuum state does.}
\end{figure}

\begin{figure}[ht]
\centering
\includegraphics[width=0.9\linewidth]{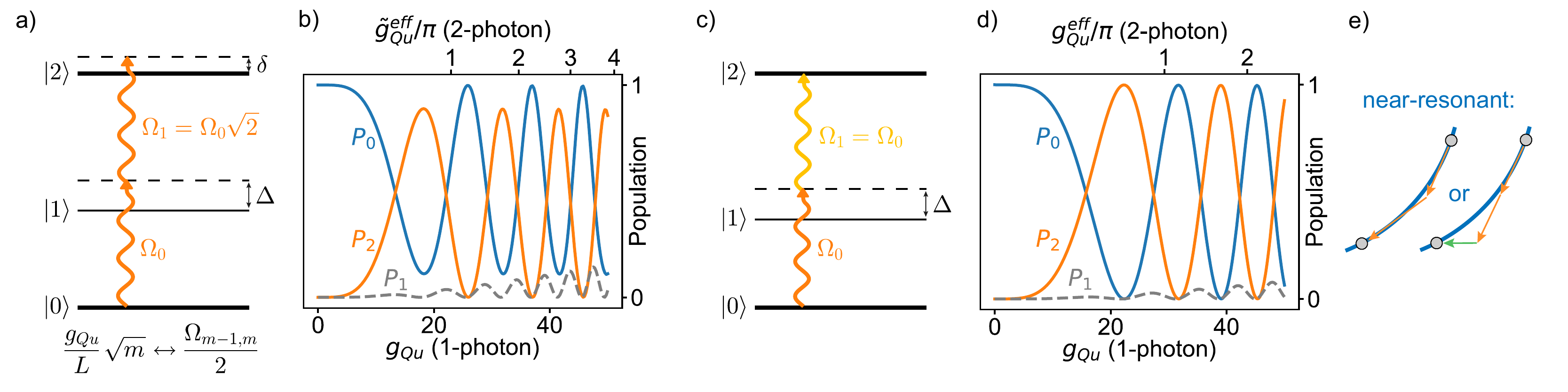}
\caption{\label{fig:3lvl}The analogy between the Raman process in a 3-level system and the electron-photon interaction when the emission of two photons is phase-matched. a-b) (single-mode case) We can approximate the electron-photon system at low $\sigma$ as the perfect 3-level system with the Rabi frequencies $\Omega_0$ and $\Omega_1 = \Omega_0\sqrt{2}$ (a). Difference in Rabi frequencies results in the effective detuning $\delta = \frac{\Omega_0^2}{4\Delta}$, which lowers the contrast of the two-photon Rabi oscillations $P_2^{max}/P_0^{max}=\frac{8}{9}\approx 0.89$ (b), similarly to the Autler-Townes effect. c-d) (two-mode case) In the two-mode case $\Omega_1 = \Omega_0$, and we see no effective detuning, so the two-photon Rabi oscillations (d) are full and the deterministic generation of photon pair is possible. e) Low-energy electrons allow the near-resonant situation, when the one-photon mismatch is small, though the system has still effectively 3~levels. Here $g_{\rm Qu}^{\mathrm{eff}} \text{(2-photon)} \equiv \frac{g_{\rm Qu}^2}{-\varphi_{-1}L}$ and $\tilde{g}_{Qu}^{eff}=g_{\rm Qu}^{\mathrm{eff}}\cdot 3/2$ (due to the effective detuning). The revivals of $P_0$ are at ${g}_{Qu}^{eff} \approx \pi$.}
\end{figure}

\clearpage
\subsection{PINEM with recoiled electrons}
\begin{figure}[ht]
\centering
\includegraphics[width=1\linewidth]{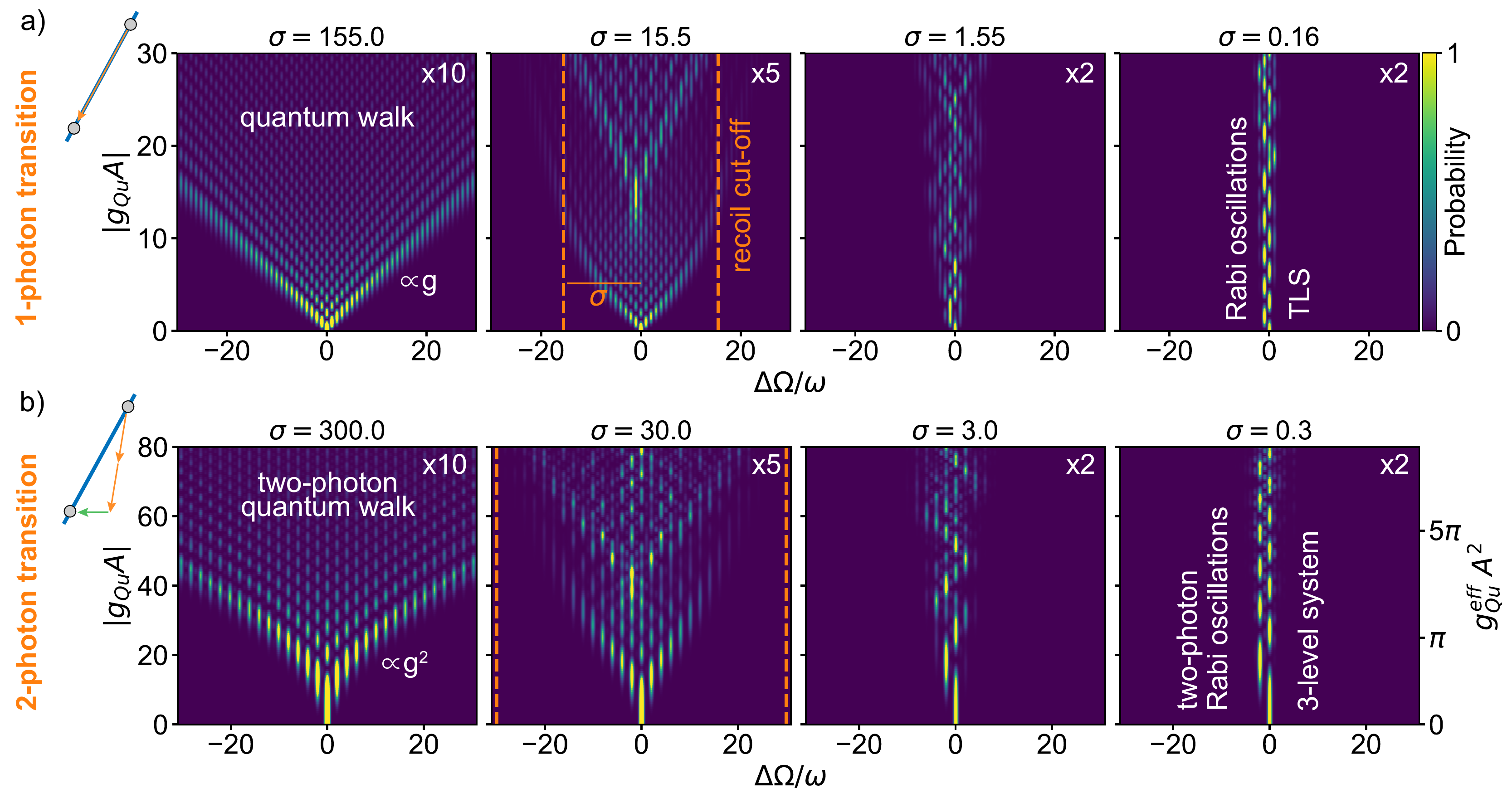}
\caption{\label{fig:pinem} Stimulated processes (PINEM). One-photon (a) and two-photon (b) PINEM with recoiled electrons showing the transition from the quantum walk to restricted PINEM with revivals and to Rabi oscillations in the two-level system (a) and to two-photon Rabi oscillations in the three-level system (b). The asymmetry of the spectra is due to the fact that the \textit{emission} (not \textit{absorption}) of one (a) and two (b) photons, respectively, is in ideal phase-matching. The effective coupling $g \propto g_{\rm Qu} A$, where $A$ is the field vector potential.}
\end{figure}

The developed approach can be also applied for the PINEM with the classical electromagnetic field to describe the electron spectrum after the interaction. We approximate that the classical field stays undepleted, so the electron and photon states are disentangled (this is similar to the parametric approximation in PDC \cite{klyshko_photons_2017}). In this case $\hat{a}^{\dagger}_{\varkappa}(t) \rightarrow A_{\varkappa} (t)$ in Eq.\ref{eq:hamilt_initial_discrete_k}, and Eq.\ref{eq:autonomized} modifies to
\begin{equation}
\begin{aligned}
f_{m}' = i\varphi_m f_{m} + (\tilde{g}_{Qu}A) f_{m-1} - (\tilde{g}_{Qu}A)^* f_{m+1}, \ \ (m = 0, \pm 1, \pm 2,..)
  \label{eq:pinem}
\end{aligned}
\end{equation}
with the "classical" coupling $\tilde{g}=(\tilde{g}_{Qu}A)=g/L$ proportional to the field strength (in agreement with Eq.\ref{eq:autonomized} for $n\gg1$ and with literature \cite{kfir_entanglements_2019, feist_quantum_2015}). Here $C_{m}(L)\equiv f_{m}(L) e^{-i\varphi_m L}$ describes the probability amplitudes of the electron side-bands. When $\sigma$ is big enough, we observe the expected quantum walk with Bessel distribution of side-bands (Extended Data Fig.~\ref{fig:pinem}). The recoil restricts the width of the electron spectrum and causes revivals of population (Extended Data Fig.~\ref{fig:phase-matching}b). It's also possible to achieve the effective two-level system (TLS) and Rabi oscillations when $\sigma<1$ (Extended Data Fig.~\ref{fig:pinem}) \cite{pan_low-energy_2023, eldar_self-trapping_2024}. 

The same considerations also describe two-photon PINEM (Fig.~\ref{fig:pinem}b). Note the quadratic spectral width growth as a function of coupling (compared to the linear dependence for one-photon PINEM), and characteristic two-photon Rabi oscillations with the zero-level revivals at $g_{\rm Qu}^{\mathrm{eff}} A^2 \approx m\pi$, $m \in \mathbb{Z}$. Thus, for example, one can use visible light as a pump in the transparency window of the cavity material, but at the same time create PINEM sidebands with the energy separation corresponding to the UV light. Note that using single-frequency pump requires QPM for the two-photon interaction with fast electrons (to mismatch the one-photon transition), though biharmonic pump can be used both with the direct phase-matching and QPM.

As a result of recoiled PINEM, the electron becomes an effective two- or three-level system, driven by a strong classical field. These situations do not require strong vacuum coupling $g_{\rm Qu}$, because it can be compensated by the intensity of the pump field ($g=g_{\rm Qu} A$). Note that with increasing $g$, the two- and three-level systems are blurred (Fig.~\ref{fig:pinem}(a-b), right panels), since mismatched processes become more probable.